\documentclass[iop]{emulateapj}

\usepackage{graphicx,times,natbib}
\usepackage{amssymb}
\usepackage{subfigure,helvet}
\usepackage{epsfig,epstopdf,soul}
\usepackage[usenames,dvipsnames]{color}
\usepackage{amsmath} 
\usepackage[export]{adjustbox}
\usepackage[usenames,dvipsnames]{color}
\usepackage{amsmath} 

\shorttitle{SF Suppression in NGC\,1266}
\shortauthors{K. Alatalo et al.}
\begin{document}

\title{Suppression of star formation in NGC\,1266}
\slugcomment{Accepted for publication to the Astrophysical Journal}


\author{Katherine Alatalo$^1$, Mark Lacy$^2$, Lauranne Lanz$^1$, Theodoros Bitsakis$^{1,3}$, Philip~N. Appleton$^1$, Kristina Nyland$^{4,5}$, Sabrina~L. Cales$^{6,7}$, Philip Chang$^8$, Timothy~A. Davis$^9$, P.~T. de Zeeuw$^{9,10}$, Carol~J. Lonsdale$^2$, Sergio Mart\'in$^{11}$, David~S. Meier$^{4,5}$ \& Patrick~M. Ogle$^1$}

\affil{
$^1$Infrared Processing and Analysis Center, California Institute of Technology, Pasadena, California 91125, USA\\
$^2$National Radio Astronomy Observatory, 520 Edgemont Road, Charlottesville, VA 22903, USA\\
$^3$Instituto de Astronom\'ia, Universidad Nacional Aut\'onoma de M\'exico, Aptdo. Postal 70-264, 04510, M\'exico, D.F., Mexico\\
$^4$Physics Department, New Mexico Tech, Socorro, NM 87801, USA\\
$^5$National Radio Astronomy Observatory, 1003 Lopezville Road, Socorro, NM 87801, USA\\
$^6$Department of Astronomy, Faculty of Physical and Mathematical Sciences, Universidad de Concepci\'{o}n, Casilla 160-C, Concepci\'{o}n, Chile\\
$^7$Department of Astronomy,Yale University, New Haven, CT 06511 USA\\
$^8$Department of Physics, University of Wisconsin - Milwaukee, Milwaukee, WI 53201, USA\\
$^9$European Southern Observatory, Karl-Schwarzschild-Str. 2, 85748 Garching, Germany\\
$^{10}$Sterrewacht Leiden, Leiden University, Postbus 9513, 2300 RA Leiden, the Netherlands\\
$^{11}$Institut de Radioastronomie Millim\'etrique, 300 rue de la Piscine, Domaine Universitaire 38406 Saint Martin d'H\`eres, France
}
\email{email: kalatalo@ipac.caltech.edu}

\begin{abstract}
NGC\,1266 is a nearby lenticular galaxy that harbors a massive outflow of molecular gas powered by the mechanical energy of an active galactic nucleus (AGN). It has been speculated that such outflows hinder star formation (SF) in their host galaxies, providing a form of feedback to the process of galaxy formation.  Previous studies, however, indicated that only jets from extremely rare, high power quasars or radio galaxies could impart significant feedback on their hosts.  Here we present detailed observations of the gas and dust continuum of NGC\,1266 at millimeter wavelengths. Our observations show that molecular gas is being driven out of the nuclear region at $\dot{M}_{\rm out} \approx 110~M_\odot$~yr$^{-1}$, of which the vast majority cannot escape the nucleus.  Only 2~$M_\odot$~yr$^{-1}$ is actually capable of escaping the galaxy. Most of the molecular gas that remains is very inefficient at forming stars.  The far-infrared emission is dominated by an ultra-compact ($\lesssim50$\,pc) source that could either be powered by an AGN or by an ultra-compact starburst. The ratio of the SF surface density ($\Sigma_{\rm SFR}$) to the gas surface density ($\Sigma_{\rm H_2}$) indicates that SF is suppressed by a factor of $\approx 50$ compared to normal star-forming galaxies if all gas is forming stars, and $\approx$150 for the outskirt (98\%) dense molecular gas if the central region is is powered by an ultra-compact starburst. The AGN-driven bulk outflow could account for this extreme suppression by hindering the fragmentation and gravitational collapse necessary to form stars through a process of turbulent injection. This result suggests that even relatively common, low-power AGNs are able to alter the evolution of their host galaxies as their black holes grow onto the M-$\sigma$ relation.
\end{abstract}

\section{Introduction}
The presence of an active galactic nucleus (AGN) has a significant effect on the growth of its host galaxy. Feedback by AGN jets or winds is thought to prevent excessive star formation (SF) in the host \citep{fabian12}. At the same time, the fuel supply to the supermassive black hole is limited by these same winds clearing gas from the vicinity of the AGN and ensuring that the central bulge components of galaxies and their black holes grow in tandem to produce a correlation between black hole mass and the velocity dispersion of their bulges \citep{ferrarese+merritt00,gebhardt+00}. Although some powerful radio galaxies and quasars show outflows that are strong enough to expel the interstellar medium (ISM) in their hosts \citep{villar-martin+99,fu+stockton09,nesvadba+10,shih+13,liu+13}, such outflows cannot be maintained for more than a few tens of millions of years, and gas will eventually accrete onto the host galaxies and form stars. A different mechanism, involving AGNs of lower luminosity, is needed to fully suppress SF in galaxy bulges on timescales comparable to the age of the Universe \citep{croton+06}, and explain the wide range of mass scales over which black hole and bulge masses are correlated.

\begin{figure*}
\includegraphics[width=\textwidth]{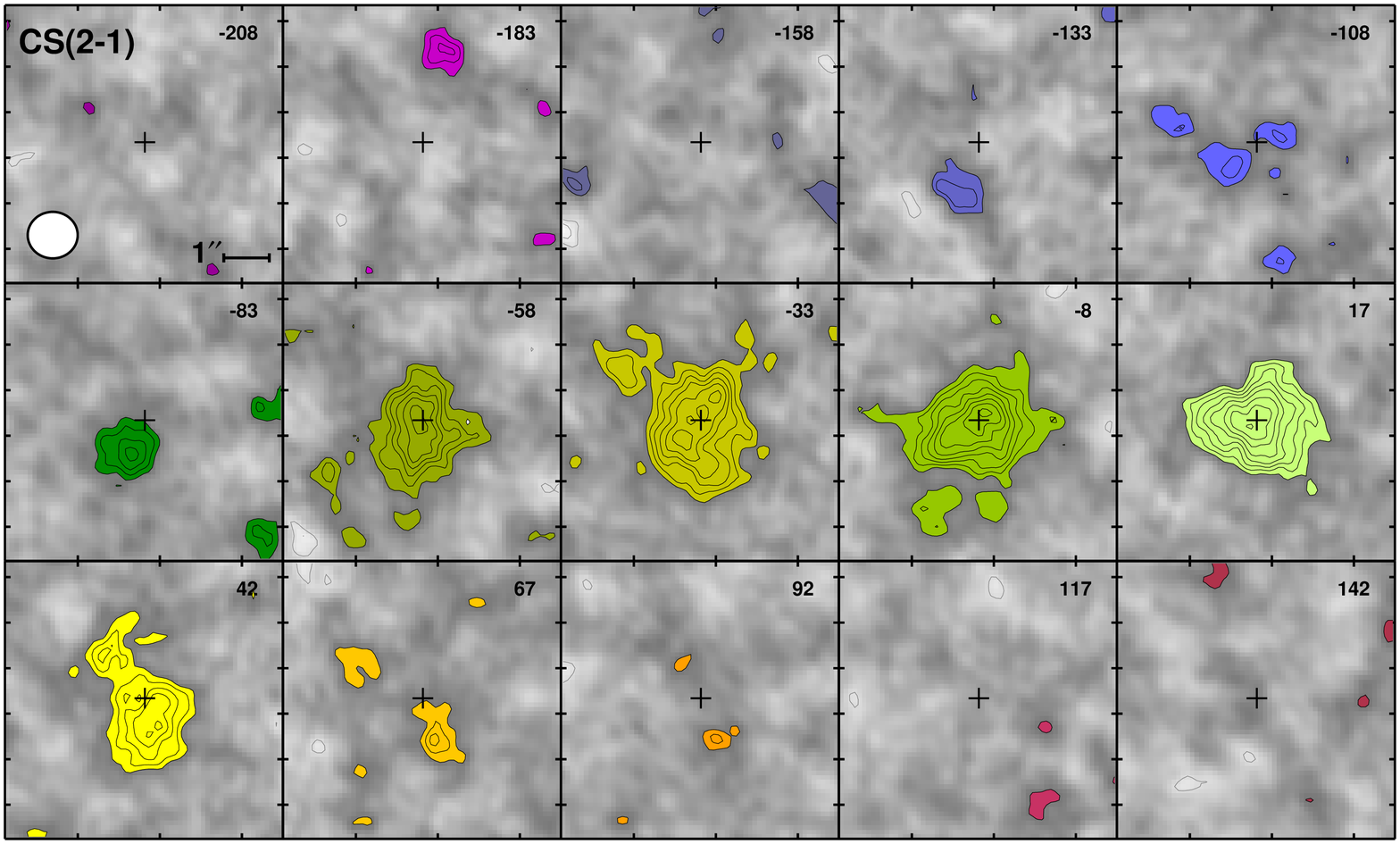}
\caption{Channel maps for CS(2--1) from CARMA.  The velocities are listed in the top corner of each panel are relative to the systemic velocity, of 2160~km~s$^{-1}$.  The first panel also shows the CARMA beam in the bottom left corner.  Contours on the plot start at $\pm3\sigma$, where $\sigma$ is the root mean square noise of the map, and are in 1$\sigma$ increments.  At the center of the line, the detections are 7--10$\sigma$ significance, with robust ($>4\sigma$) detections from -183--92~km~s$^{-1}$.  The CARMA beam (for a naturally weighted dataset) is $\approx 1''$, and it is clear that the CS emission is concentrated in the nucleus.}
\label{fig:cs_chans}
\end{figure*}

\begin{figure}
\includegraphics[width=0.49\textwidth]{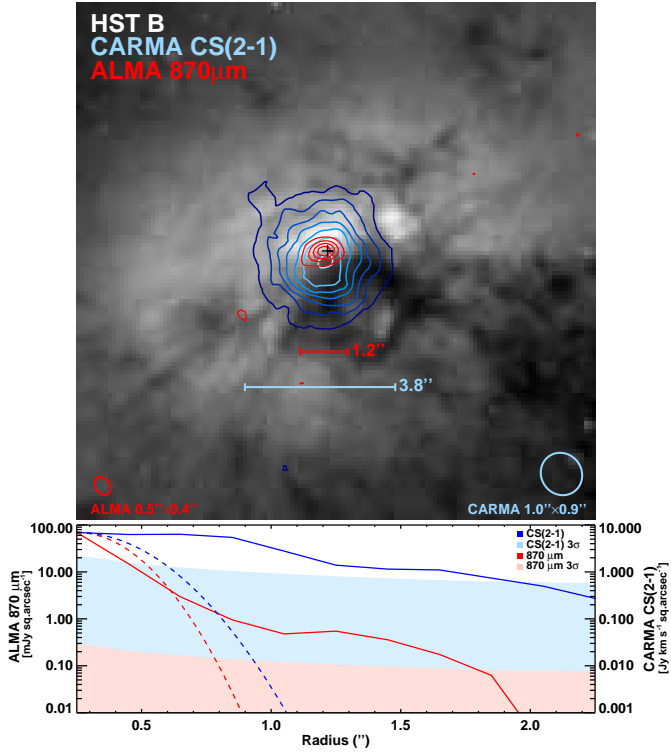}
\caption{{\bf (Top):}  CARMA CS(2--1) integrated intensity map (bluescale) contours, along with the $870\mu$m ALMA continuum data (red) contours, are overlaid atop the $B$-band HST data (grayscale; \citealt{nyland+13})  The black cross represents the VLBA point source.  CS(2--1) integrated intensity contour levels are [0.1, 0.24, 0.39, 0.53, 0.67, 0.81, 0.96] of the peak and ALMA 870$\mu$m are [3, 6, 9, 12] times the root mean square ($\sigma_{\rm rms}$) of the map.  The approximate diameter of the CS(2--1) data is 3.8$''$, equivalent to 550\,pc at the adopted distance of NGC\,1266.  The 870$\mu$m data is resolved at approximately $0.4''$, equivalent to 60\,pc (when deconvolved from the $0.5^{''}\times 0.4^{''}$ uniformly-weighted ALMA beam).  {\bf (Bottom):}  Radial profiles of the emission: 870$\mu$m (naturally weighted, $0.5^{''}\times 0.5^{''}$ beam; solid red line) and CS(2--1) (uniformly-weighted; $\theta_{\rm beam} = 0.6''\times0.5''$; solid blue line). The dashed lines show the respective beam profiles. Finally, the level of the $3\sigma_{\rm rms}$ rms noise annuli for the CS(2--1) and $870\mu$m continuum sources are shaded (light blue and light red, respectively), showing where the radial emission falls below the $3\sigma_{\rm rms}$ noise level for each respective map.  The CS(2--1) and 870$\mu$m sources both truncate at approximately the same radius $\approx 2''$, consistent with the full-width at zero intensity level of the CS(2--1) emission line image.
}
\label{fig:cs+alma}
\end{figure}

\begin{figure*}
\includegraphics[width=\textwidth]{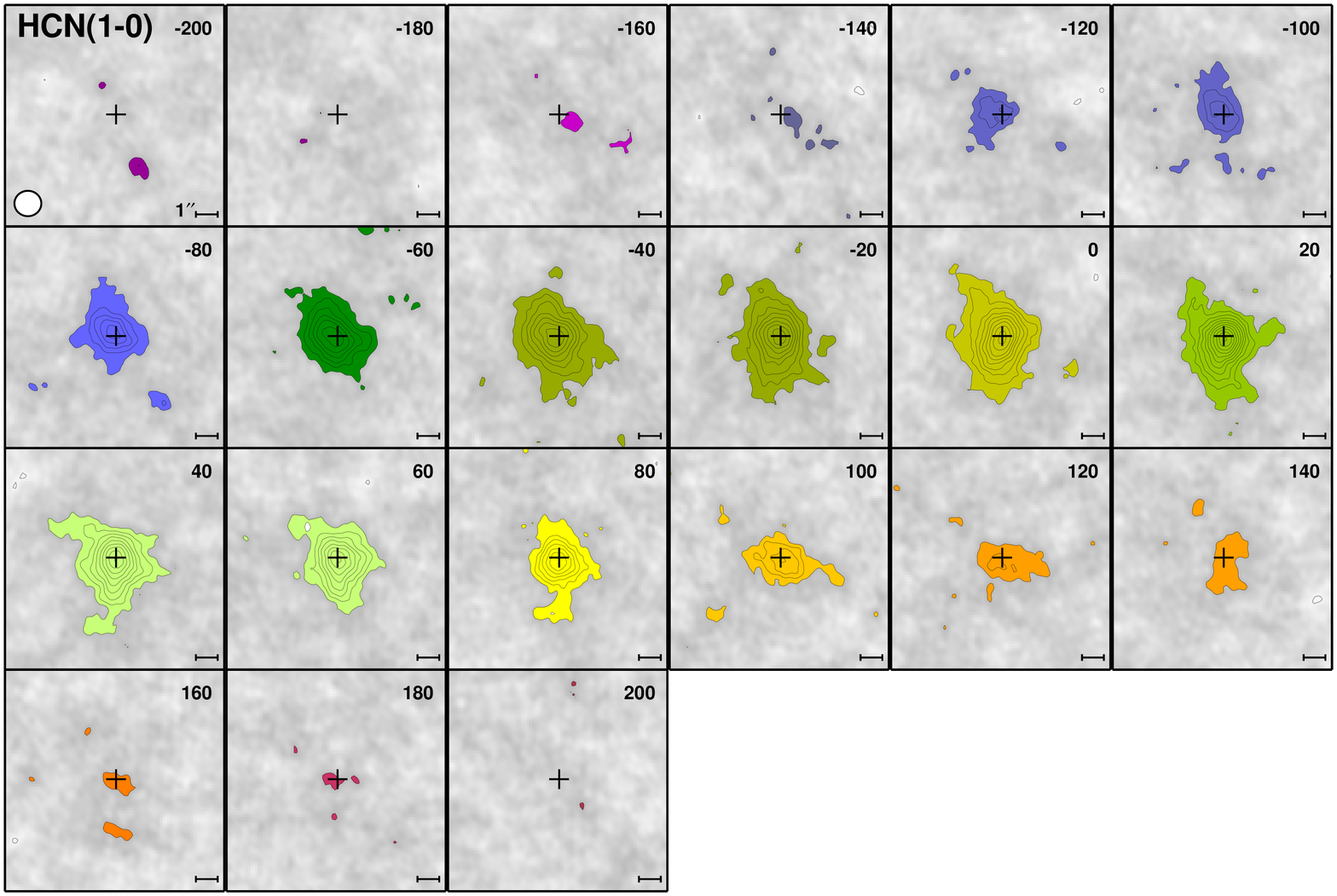}
\caption{Channel maps for HCN(1--0) from CARMA.  The velocities are listed in the top corner of each panel are relative to the systemic velocity, of 2160~km~k$^{-1}$.  The first panel also shows the CARMA beam in the bottom left corner.  Contours on the plot start at $\pm3\sigma$, where $\sigma$ is the root mean square noise of the map, and are in 2$\sigma$ increments.  At the center of the line, the detections are 7--10$\sigma$ significance, with robust ($>3\sigma$) detections from -160--180~km~s$^{-1}$.  The CARMA beam (for a naturally weighted dataset) is $\approx 1''$, and it is clear that the HCN emission is concentrated in the nucleus.}
\label{fig:hcn_chans}
\end{figure*}

Mass-loaded olecular outflows have now been seen in many AGN \citep{fischer+10,feruglio+10,sturm+11,alatalo+11,aalto+12,veilleux+13,cicone+13,gonzalez-alfonso+14}, but the mechanism by which outflows suppress SF in their host galaxies is unclear. Expulsion of molecular gas cannot be the dominant mechanism, as these outflows are usually too weak to completely remove more than a small fraction of the molecular gas from the potential well of the galaxy.  In order to make a detailed study the of suppression of SF in an AGN of moderate luminosity, we used the Atacama Large Millimeter Array (ALMA) and the Combined Array for Research in Millimeter Astronomy (CARMA) to make sensitive, sub-arcsecond resolution (corresponding to physical scales $<100$pc in both continuum and dense gas images) images of the nucleus of NGC\,1266, a nearby lenticular galaxy that harbors a massive outflow of molecular gas powered by an AGN \citep{alatalo+11}. CARMA was used to map the J=2--1 transition of carbon monosulphide (CS) and J=1--0 transition of hydrogen cyanide (HCN).  ALMA was used to map the continuum at 870$\mu$m, and to obtain continuum flux densities at lower frequencies.

Here we report on new ALMA and CARMA observations of NGC\,1266.  The distance to NGC\,1266 is taken from ATLAS$^{\rm 3D}$ \citep{cappellari+11} to be 29.9 Mpc, for which 1\arcsec\ = 145 pc.  In \S\ref{obs}, we describe the molecular line and continuum observations and analysis from ALMA and CARMA, as well as {\em Chandra} and {\em XMM-Newton}. In \S\ref{sec:sf}, we discuss the constraints on SF rate (SFR) available from the observations.   In \S\ref{sec:massoutflow} we provide new estimates of $\dot{M}_{\rm out}$, from the dense gas estimates. In \S\ref{sec:scenarios} we describe the possible energetics of the central region and discuss the implications of SF in \S\ref{sec:sfsupp}. We summarize our main conclusions in \S\ref{conc}.

\section{Observations and Analysis}
\label{obs}
\subsection{CARMA observations}

The J=2--1 transition of CS was observed in NGC\,1266 using the CARMA between December 2009 and January 2013 in the A ($\theta_{\rm beam} = 0.3''$), B ($\theta_{\rm beam} = 0.5''$), C ($\theta_{\rm beam} = 1''$), and D ($\theta_{\rm beam} = 5''$) arrays.  The spatial resolution in the final map is $1.0''\times0.9''$ and the sensitivity 1.32~mJy/beam per 20~km~s$^{-1}$ channel.  The data were reduced as described in \citet{alatalo+13}.  CS(2--1) channel maps from CARMA can be seen in Figure \ref{fig:cs_chans}.  Figure \ref{fig:cs+alma} show the integrated CS data, indicating that the total deconvolved extent of the dense gas is 3.8$''$, or 550\,pc.  

\begin{figure}
\includegraphics[width=0.47\textwidth]{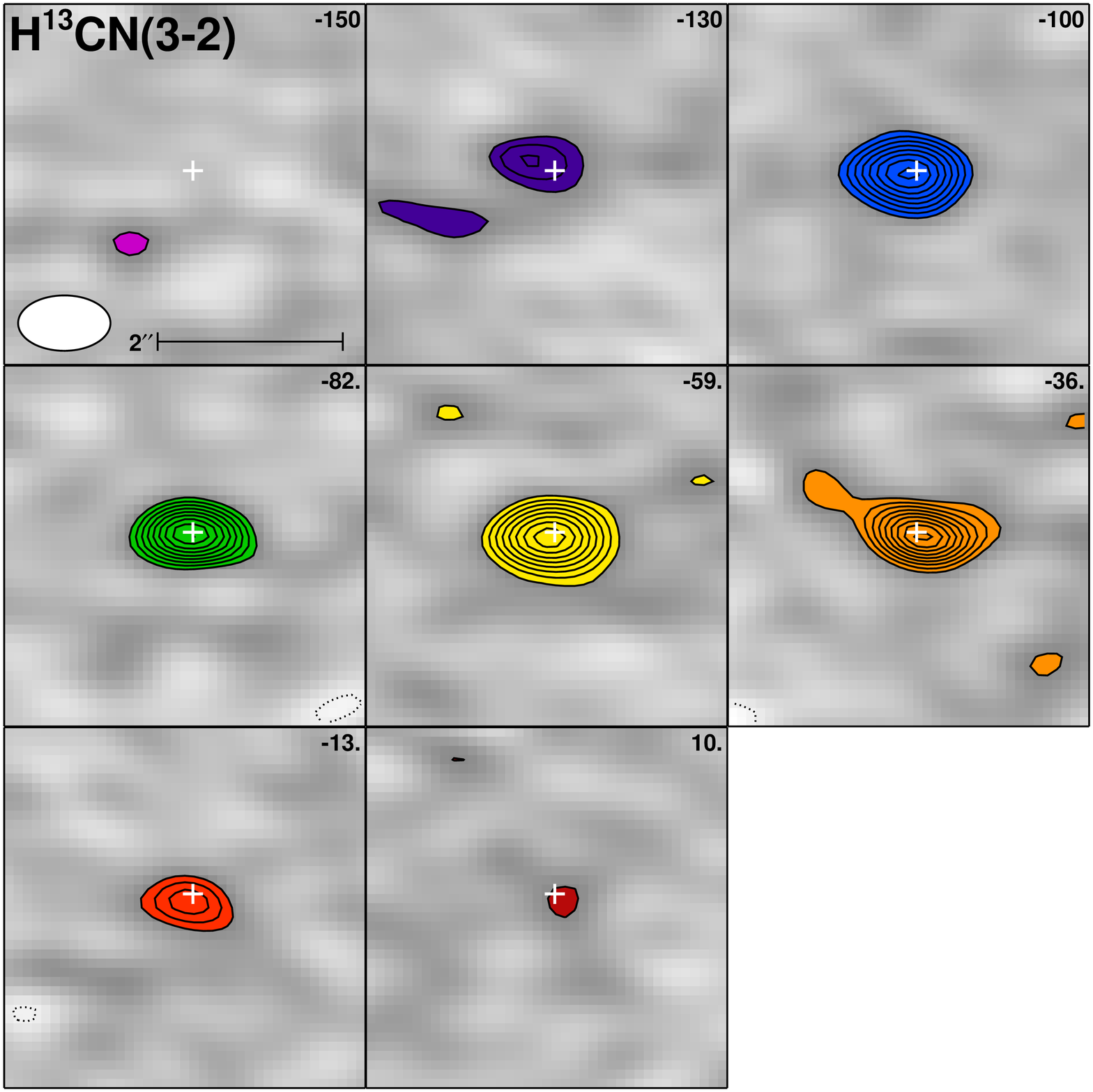}
\caption{Channel maps for H$^{13}$CN(3--2) from ALMA.  The velocities are listed in the top corner of each panel are relative to the systemic velocity, of 2160~km~k$^{-1}$.  The first panel also shows the ALMA beam in the bottom left corner.  Contours on the plot start at $\pm3\sigma$, where $\sigma$ is the root mean square noise of the map, and are in 1$\sigma$ increments, and the white cross shows the position of the VLBA point source.  The ALMA beam is $\approx 0.9'$, and it is clear that the H$^{13}$CN(3--2) emission is concentrated in the nucleus.  Deconvolving this line shows that it appears to originate and rotate about a region with radius $R<22$pc.}
\label{fig:h13cn}
\end{figure}

In order to derive the surface density of the molecular gas, we assumed that the CS gas is optically thin and in local thermal equilibrium (LTE) with an excitation temperature of 90\,K (similar to the excitation temperature of Arp\,220, and corroborated by the Carbon Monoxide (CO) spectral line energy distribution presented in \citealt{pellegrini+13}).  These assumptions and the CS abundance of $\approx10^{-9}$ lead to a column density of $N({\rm H_2}) \approx 1.05\times10^4$~M$_\odot$~pc$^{-1}$.  The gas column was also calculated to be $1.13\times10^4$~M$_\odot$~pc$^{-1}$ using the CS-to-$N({\rm H_2})$ relation found in infrared (IR) dark clouds in the Milky Way \citep{lada+94} of \hbox{$I$(CS J=2--1)/(K km s$^{-1}$) = 0.10 + 0.06 mag$^{-1}~A_V$} and \hbox{$N(\rm H_2)/A_V = 9\times10^{20}$ cm$^{-2}$ mag$^{-1}$} from \citet{schultz+75}.  From these methods, the derived average molecular gas surface density is $\Sigma_{\rm H_2} \approx 10^4$ M$_\odot$~pc$^{-2}$, and includes all points in the CS(2--1) moment0 map within a radius of 1.9$''$.  This is likely a lower limit, given the uncertainties in the CS abundance and opacity.  The derived $\Sigma_{\rm H_2}$ agrees reasonably well with the CO(1--0) derived $\Sigma_{\rm H_2}$ of $2.7\times10^4$ M$_\odot$ pc$^{-2}$ \citep{alatalo+11}, suggesting a significant fraction of the total molecular gas is dense.  The line-of-sight column to the Very Long Baseline Array (VLBA) radio point source \citep{nyland+13} is found to be N$_{\rm H_2,dense} \approx 3\times10^{24}$~cm$^{-2}$.  

The HCN(1--0) data were also taken at CARMA and reduced and analysed identically to the CS(2--1).  In order to determine the total mass and therefore gas surface density associated with the HCN, we assumed a HCN to H$_2$ abundance of 10$^{-8}$ and LTE with an excitation temperature of 90\,K.  The inferred HCN mass was $2 \times10^9$~M$_\odot$, which corresponded to a gas surface density of $0.8\times10^4$ in the dense gas region, similar to what was derived for CS.  Given that both the HCN(1--0) and CS(2--1) trace only the dense gas and agree that $\Sigma_{\rm H_2} \approx10^4$~M$_\odot$~pc$^{-2}$, it is likely that the CS-derived quantities provide a physically robust picture of the dense gas properties of NGC\,1266.   HCN(1--0) channel maps can be seen in Figure \ref{fig:hcn_chans}.  It is of note that the CS(2--1) and HCN(1--0) maps appear to show disturbed molecular gas.

\subsection{ALMA observations}
\label{sec:almaobs}

The ALMA observations of NGC\,1266 were made in the extended Cycle 0 configuration as part of project 2011.0.00511.S, and processed through the standard ALMA Cycle 0 reduction CASA\footnote{http://casa.nrao.edu/} script by ALMA staff. We used the calibrated measurement set product to image the continuum at 344.7~GHz ($870\mu$m), excluding any bright molecular lines. A single iteration of self-calibration in phase only was applied to the dataset.  Both naturally-weighted (with a $0.51^{''} \times 0.47^{''}$ beam) and uniformly-weighted (with a $0.49^{''} \times 0.36^{''}$ beam) images were made.  

H$^{13}$CN(3--2) emission at 259.0118 GHz ($\lambda_0 = 1157.45\mu$m) was detected in one of the ALMA band 6 observations, which was reduced to a calibrated measurement set by ALMA staff in the same procedure as the ALMA band 7 observations. We then imaged the datacube using natural weighting and a channel width of 2.44\,MHz (3.34\,km\,s$^{-1}$) (no self-calibration was applied). Figure \ref{fig:h13cn} shows the channel maps of H$^{13}$CN(3--2) from ALMA, showing that the dense gas strongly overlaps the radio point source. A virial equilibrium estimate was made using this line (which has a deconvolved size of $\lesssim0.25''$ and radius of $\lesssim22$pc) require an enclosed mass of $\approx4.5\times10^7$ M$_\odot$.  This virial estimate predicts a mass surface density of $\Sigma_{\rm *,\bullet,gas} \gtrsim 2.9\times10^4$~M$_\odot$~pc$^{-2}$, which supports the gas surface density derived from the CS(2--1) and HCN(1--0) observations from CARMA.

The ALMA continuum and H$^{13}$CN(3--2) maps show a barely-resolved central source $\lesssim 50$~pc in size.  Weak, diffuse emission is detected when integrating over successive annuli in the naturally-weighted map to create a radial profile of the ALMA data between $1^{\prime \prime}$ and 2$''$, consistent with the edge of the CS disk.  We mapped the $870\mu$m data using a naturally-weighted beam to maximize our sensitivity to diffuse emission. We used the CS moment0 map to define the region within which to measure the total flux density, and used the CASA {\sc imfit} task to fit the central source (allowing a zero-point offset to account for the diffuse emission). We found an integrated flux density of 32.6$\pm 0.1$ mJy up to an aperture of 4.0$''$\ diameter with the compact central source contributing 23.5$\pm 0.5$mJy, and the residual diffuse emission contributing 9.1$\pm 0.5$mJy. 

In order to test for continuum flux resolved out by ALMA on the scale of the CS(2--1) molecular gas emission, we used the {\sc simdata} and {\sc simanalyse} tasks in CASA to simulate imaging the spatial distribution of the CS emission from CARMA using the same configuration and frequency as the ALMA Cycle 0 $870\mu$m observations. The simulation recovered 70\% of the flux in the CS(2--1) image. Under the assumption that this emission is co-spatial with the diffuse continuum emission at 870$\mu$m, we corrected the ALMA extended flux component upwards by 30\% to compensate.  The spatial decomposition provided at 870$\mu$m by the ALMA data shows that the emission coming from the more extended gas disk accounts for only about 34\% of the flux after the correction is applied. Lower frequency measurements of the continuum at 994$\mu$m, 1165$\mu$m, 1381$\mu$m and 3476$\mu$m from the same ALMA program were used to construct the SED.

\subsection{XMM-Newton and Chandra Observations}
\label{sec:xrayobs}
\begin{figure*}[t]
\includegraphics[width=\textwidth]{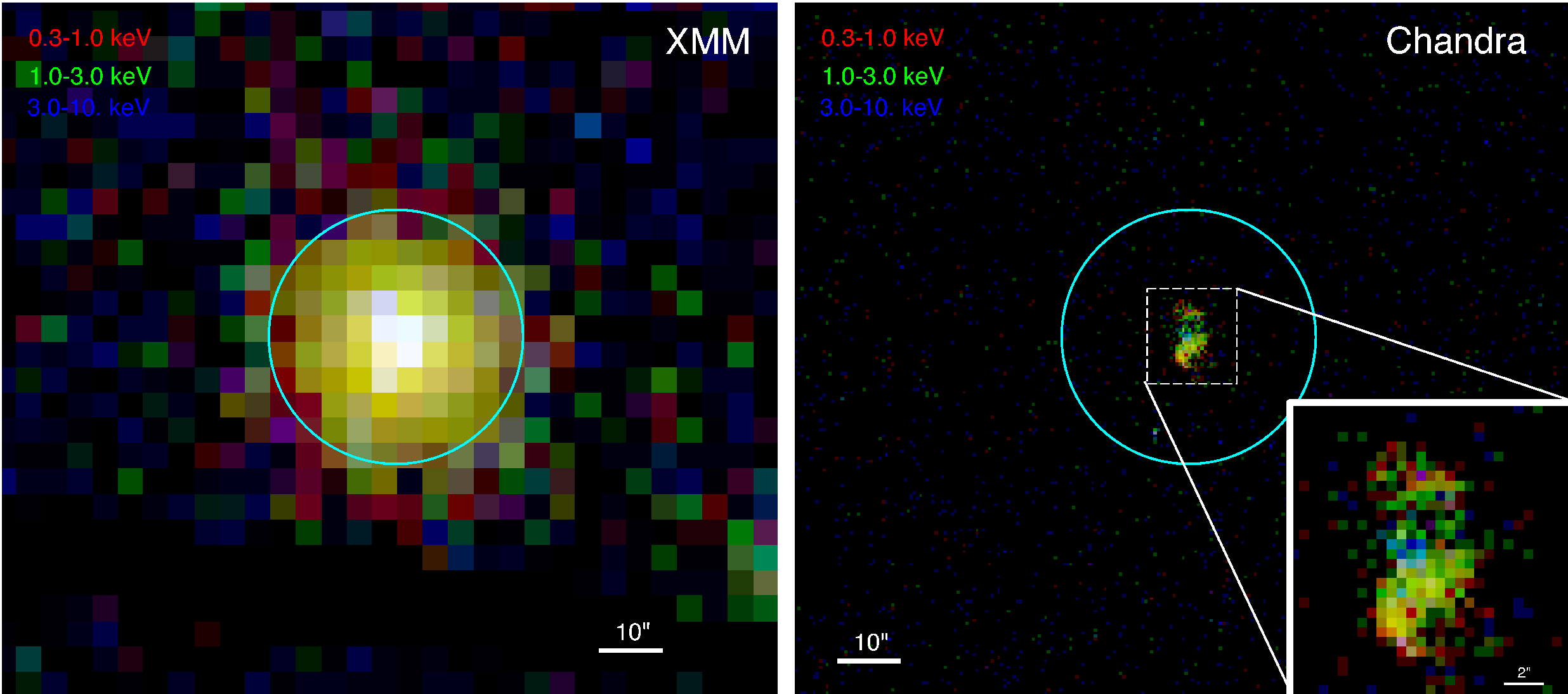}
\caption{3-color images of the reduced, unsmoothed {\em XMM} (left) and {\em Chandra} (right and inset) X-ray data.  Similar to what was presented in Figures 11 \& 12 of \citet{alatalo+11}, the {\em Chandra} images show that the bulk of the X-rays originate in the shocked outflow (consistent with the position of the outflowing ionized gas from \citealt{davis+12}).  These 3-color images also shows that the hard X-rays are concentrated in a single location, near the position of the AGN \citep{nyland+13}, agreeing well with the original conclusions of \citet{alatalo+11}. The cyan circle shows the spectral extraction region, defined based on the extent of the X-rays detected by {\em XMM}.}
\label{fig:xray_image}
\end{figure*}

{\em XMM-Newton} observed NGC\,1266 for 138.58\,ks on 2012 July 25 (ObsID 0693520101). We retrieved and analyzed the data taken with European Photon Imaging Camera \citep[EPIC; ][]{jansen01} on both the Metal Oxide Semiconductor CCDs (MOS) and pn CCDs. We filtered each dataset to remove periods of background flaring, resulting in reduced exposure times of 91.49\,ks (MOS1), 107.31\,ks (MOS2), and 91.68\,ks (pn). The events were further filtered to retain only events with energies between 0.3\,keV and 10\,keV and patterns between 0 and 12. We extracted the spectrum in a 20$''$ aperture (shown in Figure \ref{fig:xray_image}) centered on the position of the VLBA point source \citep{nyland+13} in each dataset and then combined them to create the EPIC spectrum. After filtering, this aperture contains $9190\pm100$ net counts between 0.3\,keV and 10\,keV. 

NGC\,1266 was observed for 29.65\,ks on 2009 September 20 (ObsID 11578) with the back-illuminated CCD chip, S3, of the {\em Chandra} Advanced CCD Imaging Spectrometer \citep[ACIS;][]{weisskopf00}. We reprocessed the observation using {\sc{ciao}} version 4.5 to create  a new level 2 events file, following the software threads from the \emph{Chandra} X-ray Center (CXC)\footnote{\url{http://cxc.harvard.edu/ciao}}. An X-ray spectrum was extracted using the {\sc specextract} task in the 0.4$-$8.0\,keV energy range in the same aperture as the {\em XMM} spectrum. However,  the bulk of the photons observed by \emph{Chandra} are within a 6$''$ aperture due to the better resolution of \emph{Chandra} compared to \emph{XMM} (see Fig. \ref{fig:xray_image}). This aperture contains $857\pm32$ net counts between 0.4\,keV and 8.0\,keV. We filtered the events based on that energy range and then grouped it to a minimum of 30 counts per bin prior to modeling the spectrum. 

\begin{figure}[b]
\includegraphics[width=0.48\textwidth,clip,trim=1.5cm 0.9cm 1.7cm 1.5cm]{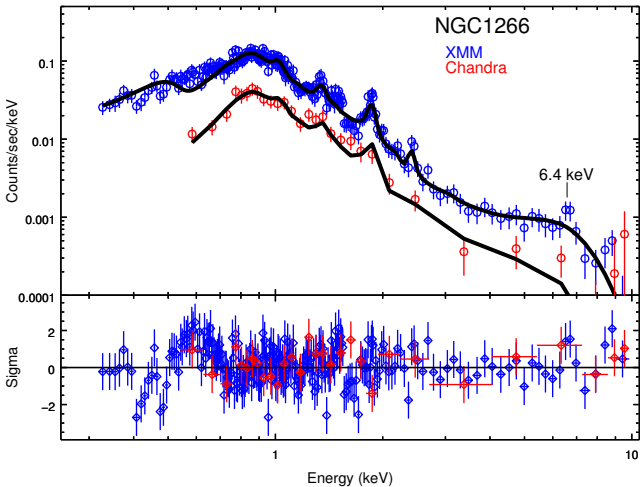}
\caption{We simultaneously fit the spectra of the {\em XMM} (blue) and {\em Chandra} (red) extracted in a 20$''$aperture (see Fig. \ref{fig:xray_image}), with an absorbed model comprised of a thermal component, a power-law component, and five lines, including Fe~K$\alpha$ (marked; with varying width), which is shown in black.  The 6.4~keV {\em XMM} bin (position of Fe~K$\alpha$) shows a boosted signal, consistent with the presence of a broad Fe~K$\alpha$ line.}
\label{fig:xray_spec}
\end{figure}

Spectral modeling was done using the {\sc sherpa} packages of {\sc ciao} using the Levenberg-Marquardt optimization method \citep{more78}. The same model was applied to both the {\em Chandra} and EPIC spectra and fit simultaneously. A fixed foreground absorption due to the Milky Way ISM of $N_{H}=5.55\times10^{20}\,{\rm cm^{-2}}$ \citep{kalberla05}\footnote{\url{http://heasarc.nasa.gov/cgi-bin/Tools/w3nh/w3nh.pl}} is assumed in addition to intrinsic absorption. Both a purely thermal model \citep[APEC; ][]{smith01} and a power law model can be ruled out ($p<<0.001$) even when intrinsic absorption is included. An absorbed combination of a thermal and power law model does better ($\chi^2$/dof=375.3/263), but the inclusion of five lines modeled as gaussians is needed to make the fit statistically good ($\chi^2$/dof=254.5/257). Each line significantly improves the fit, particularly in the {\em XMM} spectrum (Figure \ref{fig:xray_spec}). The best fit model has $kT=0.74\pm0.02$\,keV, $\Gamma=3.0\pm0.2$, $N_{H}=(1.5\pm0.2)\times10^{21}$\,cm$^{-2}$, and lines at 6.400\,keV (Fe\,K$\alpha$), 1.870\,keV (Si\,{\sc xiii}), 1.352\,keV (Mg\,{\sc xi}), 2.402\,keV (Si\,{\sc xiii}), and 1.022\,keV (Ne\,{\sc x}). Line identifications are tentative and are based on the most intense line at that energy given in {\sc atomdb}\footnote{\url{http://www.atomdb.org}}. The fit also gives Fe K$\alpha$ a width of $\sigma=1.7\pm0.5$\,keV. The thermal component has an unabsorbed 0.3$-$10.0\,keV luminosity of $8.9^{+0.7}_{-0.6}\times10^{39}$\,erg\,s$^{-1}$ and the power law has an unabsorbed 2$-$10\,keV luminosity of $2.7^{+0.46}_{-0.3}\times10^{39}$\,erg\,s$^{-1}$.  Figure \ref{fig:xray_image} shows the {\em XMM} and {\em Chandra} 3-color images, showing the relative positions and distributions of the soft and hard X-rays.  In the {\em Chandra} image, a blue point-like source appears approximately at the location of the NGC\,1266 nucleus.

The dense gas observations (discussed above) show that the line-of-sight gas column ($N_{\rm H_2}\approx3\times10^{24}$ cm$^{-2}$) towards the AGN is three orders of magnitude greater than that derived from the X-ray spectral fit ($\sim10^{21}$ cm$^{-2}$).  This could be due to the spatial complexity of the source, with multiple X-ray emitting components located within the $20''$ aperture.  The {\em Chandra} image seems to indicate this is the case, with the majority of the X-rays distributed in a similar fashion to the ionized gas from shocks \citep{alatalo+11,davis+12}, with the hard X-rays coming from a much more compact, unresolved region near the nucleus, and position of the VLBA point source \citep{alatalo+11,nyland+13}. We tried alternative spectral fits to account for this discrepancy. Separate absorbing columns on the power law (the putative AGN) and diffuse (APEC and lines) components yielded a similarly good fit to the spectra with similar parameters and requiring no intrinsic absorption on the diffuse component, but the absorbing column on the power law component still fit with $N_{H}\sim10^{21}$\,cm$^{-2}$. Fits which forced a single or one of two absorbing columns of $N_H\approx6\times10^{24}$\,cm$^{-2}$ failed, yielding very high $\chi^2/$dof values along with very unlikely values of the model parameters, often at the edge of their allowed ranges.

To determine how powerful an AGN could be hidden by the column given by the dense gas observations to yield the limited observed hard X-ray emission, we simulated the spectrum of a power law with a typical $\Gamma=2$ absorbed by a covering column $N_{H}=2N_{\rm H_2}=6\times10^{24}$~cm$^{-2}$ (We use $N_H$ rather than $N_{\rm H_2}$ here because X-ray absorption occurs via the metals; \citealt{morrison+83}). We determine the normalization of the model power law by measuring the 6--10\,keV flux of the simulated spectrum and matching it to the observed flux in this band. We use the hard X-ray band to set the normalization in order to minimize the contribution of the diffuse, thermal emission, which dominates at lower energies. The spectrum is then synthesized over the energy range of {\em Chandra} and {\em XMM}, allowing us to measure the 2--10\,keV luminosity of the putative AGN, which, once corrected for intrinsic and Galactic absorption, is $7\times10^{43}$~erg\,s$^{-1}$, which is larger than the IR luminosity in the nucleus of NGC\,1266.  Given the possibility that the hard X-rays detected by {\em Chandra} and {\em XMM} could be reflected into the line-of-sight through preferentially less obscured lanes \citep{antonucci+85,urry+95,levenson+06}, this X-ray luminosity defines the upper limit.

The IR observations (discussed Section \ref{sec:scenario1}) indicate that the bolometric luminosity of the central source is $3.4\times10^{43}$\,erg\,s$^{-1}$, smaller than the 2--10\,keV luminosity our simulated spectrum. Therefore, we simulated a second set of spectra whose underlying power law ($\Gamma=2$) has a 2--10\,keV luminosity set by the bolometric luminosity (assuming the L$_{X}$/L$_{\rm bol}=28$ ratio of Seyferts from \citealt{ho2008}). We simulated spectra with a range of covering columns up to $6\times10^{24}$\,cm$^{-2}$ to determine the $N_{H}$ that would produce the observed 6--10\,keV flux. For an expected L$_{\rm 2-10\,keV}\approx10^{42}$\,erg\,s$^{-1}$, we find the observed hard X-ray flux requires a covering column of $N_{H}=3\times10^{24}$\,cm$^{-2}$ (or $N_{\rm H_2} \approx 1.5\times10^{24}$\,cm$^{-2}$, which is comfortably within the range of that predicted by the molecular gas.

\section{Constraints on the star formation rate}
\label{sec:sf}
\begin{figure*}[t!]
\includegraphics[width=6.5in,clip,trim=1cm 0.5cm 0cm 0cm]{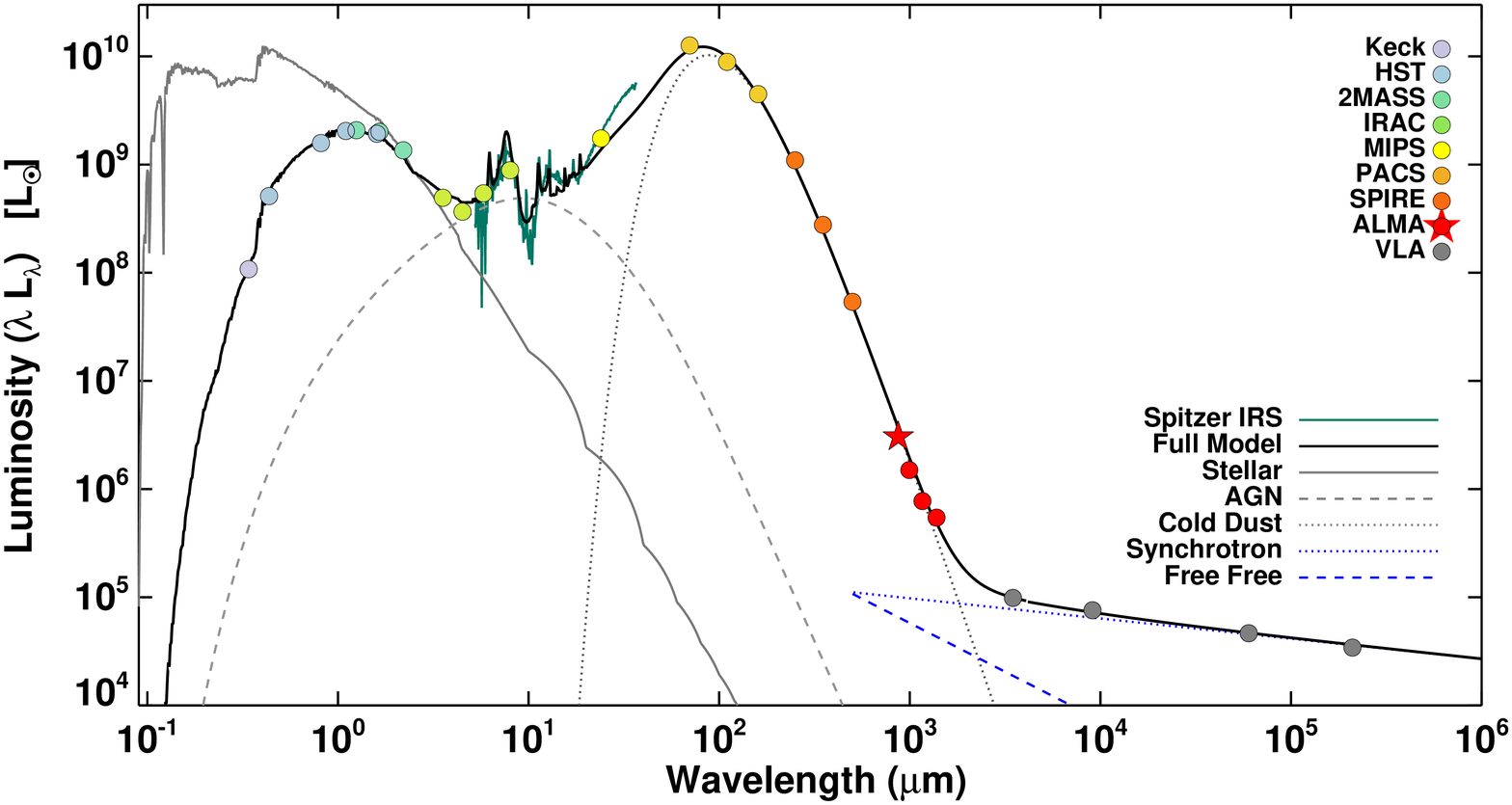}
\caption{A full SED fit to NGC\,1266 within a 15$''$ diameter aperture using the models of \citep{sajina+06}, which fit SEDs of star-forming regions in deeply embedded gas disks, and includes stellar emission, AGN emission,  and blackbody dust emission.  The SED points shown on the plot are derived within the aperture and range from $u'$-band and to VLA 21cm continuum, including Keck $u'$ (this work), HST \citep{nyland+13,alatalo+14}, 2MASS \citep{2mass}, {\em Spitzer} \citep{dale+05}, {\em Herschel} \citep{dale+12}, ALMA (this work) and VLA continuum \citep{nyland+13}.  The 870$\mu$m ALMA point is demarcated with a star.  The {\em Spitzer} IR Spectrograph data \citep{dale+06} are overplotted (green line) to further aid in constraining the mid-IR contribution.  Calibration uncertainty error bars are included, but sufficiently small that their extent is within the points on the plot.}
\label{fig:n1266_sed}
\end{figure*}

The SFR in NGC\,1266 is not straightforward to derive. The nuclear region is both highly obscured and contains large amounts of shocked gas \citep{davis+12}, resulting in important caveats on traditional SFR estimators. If there is a buried AGN in NGC\,1266, given the deep column of gas sitting directly along the line-of-sight, any radiatively powerful AGN would be Compton-thick, and also have a high optical depth even to mid-IR emission. The AGN therefore would make an unknown contribution to the total IR flux, causing an overestimate of the SFR. \citet{davis+12} showed that the ionized gas emission line ratios in the system are consistent with shocks, evidenced not only from the line diagnostic diagrams \citep{bpt,vo87,kewley+06}, but also from the large-scale velocity structures.  This would therefore contaminate attempts to determine the SFR using ionized gas lines from H~{\sc ii} regions.  In the following, we use the available tracers of SF to define lower and upper limits to the total SF taking place in the dense gas in NGC\,1266.  The 24$\mu$m flux will also overestimate the SFR, as it is clear that the AGN in the system contributes non-negligibly to the mid-IR flux (Fig. \ref{fig:n1266_sed}).

\subsection{The lower limit to SF: PAH+UV estimate}
In order to derive a lower limit to the SFR in NGC\,1266, we use the far-Ultraviolet (UV) emission from the Galaxy Evolution Explorer (GALEX; \citealt{galex}), combined with the SFR derived from the Polycyclic Aromatic Hydrocarbon (PAH) emission.  Given the extreme extinction in the far-UV, as well as the possible PAH destruction by shocks in the system, we consider this estimate to be a lower limit to the total SFR. UV photons from young stars form the basis of most techniques to estimate SFRs; typically, however, in a dense, dusty object like the nucleus of NGC\,1266, they are reprocessed into IR emission. Ionizing photons from the AGN also add to the UV emission, however. We can obtain a firm lower limit on the SFR from the UV photons that do directly escape the star-forming region and appear in the GALEX far-UV image, this is 0.003\,$M_{\odot}{\rm yr^{-1}}$ \citep{leroy+08,alatalo+11}.  We combine this SFR with that derived using a fit to the PAH emission. These large molecules are excited by UV photons and re-emit in the mid-IR. The {\em Spitzer} Infrared Spectrograph (IRS; \citealt{spitzer_irs}) spectrum was fitted using PAHFIT \citep{smith+07}, and the derived SFR from the PAH emission is 0.3 M$_\odot$~yr$^{-1}$ from \citet{treyer+10}, normalized to a Salpeter Initial Mass Function (IMF; \citealt{salpeter55}).  Due to the large levels of extinction toward the SF region and the possibility that shocks have destroyed some PAHs, we consider the SFR derived from the PAH+far-UV emission (of $\approx0.3$~M$_\odot$~yr$^{-1}$) to represent a lower limit.  It is possible that the poststarburst stellar population, as well as the AGN itself might excite the PAH emission, but compared to the shock destruction and extinction, A-star radiation is likely not the dominant effect.

\subsection{The upper limit to SF: Total IR vs. [Ne~{\sc ii}]}
Second, the UV flux absorbed from the SF and AGN and re-emitted in the far-IR can be estimated by fitting the optical-to-submillimeter Spectral Energy Distribution (SED). To create this, we used aperture photometry through a 15$^{''}$ diameter aperture subtracting the corresponding sky and estimating the flux 3$\sigma$ above the general background. Radio fluxes from the Very Large Array (VLA), described in \citet{nyland+13} were also derived using the $15''$ aperture.  The SED was fit utilizing the models of \citet{sajina+06}, which is meant to be used for deeply embedded star-forming regions (likely the case for NGC\,1266).  The fluxes derived for the SED fit can be found in Table \ref{tab:sed} and Figure \ref{fig:n1266_sed}.  Integrating the cold component of the dust and assuming a Salpeter IMF \citep{salpeter55} yields a SFR (assuming that all IR luminosity is from SF) of $\lesssim 2.2$ M$_\odot$~yr$^{-1}$.

The [Ne~{\sc ii}] and [Ne~{\sc iii}] emission from the nucleus of NGC\,1266 \citep{dudik+09} indicates a SFR of $\approx 1.5$ M$_{\odot}$ yr$^{-1}$ from this region \citep{ho+keto07}.  This [Ne~{\sc ii}]-derived SFR can be thought of as an upper limit as well, as this line is almost certainly contaminated by emission from the shocks, which is the dominant energising mechanism of the ionised gas \citep{davis+12}.


\begin{table}[b]
\centering
\caption{SED values for NGC\,1266 in a 15$''$ aperture}
\begin{center}
\begin{tabular}{lllc}
\hline \hline
Telescope & Filter & $\lambda$ & Flux \\
& Band & ($\mu$m) & (10$^{-14}$ W m$^{-2}$)\\
\hline
Keck & {\em u}$'$ &  \phantom{00000}0.340 &  0.397$\pm$0.009\\
{\em HST} &  {\em B} &   \phantom{00000}0.435 &      1.88$\pm$0.10\\
{\em HST} &  {\em I} &   \phantom{00000}0.814 &      5.82$\pm$0.18\\
{\em HST} &  {\em Y} &    \phantom{00000}1.10 &      7.52$\pm0.27$\\
2MASS & {\em J} &     \phantom{00000}1.25 &      7.65$\pm$0.36\\
{\em HST} & F160W &     \phantom{00000}1.60&      7.09$\pm$0.28\\
2MASS & {\em H} &     \phantom{00000}1.65&      7.39$\pm$0.36\\
2MASS & {\em K$_s$} &     \phantom{00000}2.20&      4.99$\pm$0.20\\
{\em Spitzer} & IRAC1 &      \phantom{00000}3.55&      1.82$\pm$0.08\\
{\em Spitzer} & IRAC2 &     \phantom{00000}4.50&      1.35$\pm$0.08\\
{\em Spitzer} & IRAC3 &     \phantom{00000}5.80&      2.00$\pm$0.10\\
{\em Spitzer} & IRAC4 &     \phantom{00000}8.00&      3.26$\pm$0.21\\
{\em Spitzer} & MIPS24 &     \phantom{0000}24.0&      6.42$\pm$0.50\\
{\em Herschel} & PACS70 &     \phantom{0000}70.0&      46.5$\pm$2.1\\
{\em Herschel} & PACS110 &     \phantom{000}110 &      32.7$\pm$1.4\\
{\em Herschel} & PACS160 &     \phantom{000}160 &      16.5$\pm0.6$\\
 {\em Herschel} & SPIRE250 &    \phantom{000}250 &      4.03$\pm$0.20\\
{\em Herschel} & SPIRE350 &     \phantom{000}350 &      1.02$\pm$0.05\\
{\em Herschel} & SPIRE500 &     \phantom{000}500 &     0.20$\pm$0.01\\
ALMA & Band 7 &    \phantom{000}870 &    0.011$\pm0.001$\\
ALMA & Band 7 &  \phantom{000}994 & 0.0055$\pm$0.0005\\
ALMA & Band 6 & \phantom{00}1165 & 0.0029$\pm$0.0004\\
ALMA & Band 6 & \phantom{00}1381 & 0.0020$\pm$0.0003\\
ALMA & Band 3 & \phantom{00}3476 & 0.00036$\pm$0.00001\\
VLA & 33 GHz & \phantom{00}9090 & 0.00028$\pm$0.00001\\
VLA & C-band & \phantom{0}60000 & 0.00017$\pm$0.000005\\
VLA & L-band & 210000 & 0.00013$\pm$0.000001\\
\hline \hline
\end{tabular}
\end{center}
\label{tab:sed}
\end{table}

\subsection{Star formation rate derived from millimeter free-free emission}
Finally, we used ALMA and VLA data to constrain the free-free radio continuum emission, arising from the ionized gas in the SF regions, which we consider the most robust estimator of the SFR in NGC\,1266. Continuum emission was measured at 86.3~GHz, 217.2~GHz, 257.6~GHz, and 301.8~GHz with ALMA and 1.4~GHz, 5~GHz and 33~GHz with the VLA and fit to a model consisting of power-law synchrotron and free-free components.  The radio continuum is resolved on scales $<15''$ \citep{nyland+13}, consistent with the aperture chosen to derive the complete SED.  The high accuracy of the {\em Herschel}, ALMA and VLA flux measurements allow us to constrain the free-free emission well, despite its relatively small contribution to the overall SED. Our best fit value for the free-free flux density at 200~GHz is 0.7~mJy (compared to a synchroton flux density of 1.7~mJy and a flux density from dust emission of 4.9~mJy). The model allows a 2$\sigma$ upper limit of 0.9~mJy on the free-free emission, and also shows that a free-free contribution is required to fit the data, with a 2$\sigma$ lower limit of 0.5~mJy at 200~GHz. The main uncertainty in our derived value of the free-free emission is any curvature in the synchrotron spectrum, however, we see no evidence for that at lower frequencies. 

Using the standard relation between SFR and free-free emission (equation 6 in \citealt{murphy+12}), we derived a SFR of 0.87$\pm 0.1$ $M_\odot$~yr$^{-1}$. Although free-free emission can be reduced in very dusty H{\sc ii} regions if ionizing photons are absorbed before they are able to ionize a hydrogen atom \citep{murphy+12}, such dust grains will re-emit in the IR, so will be accounted for in our SED fit. We estimate the overall uncertainty on this estimate to be about a factor of two given possible contributions to the free-free emission from gas ionized by AGN emission or shocks, and the possibility of dust-absorption within the H~{\sc ii} regions.  The uncertainty in this estimator place the upper limit at the same value as what was derived from the [Ne~{\sc ii}].

\begin{figure}
\includegraphics[width=0.49\textwidth,clip,trim=0.5cm 0cm 0cm 0cm]{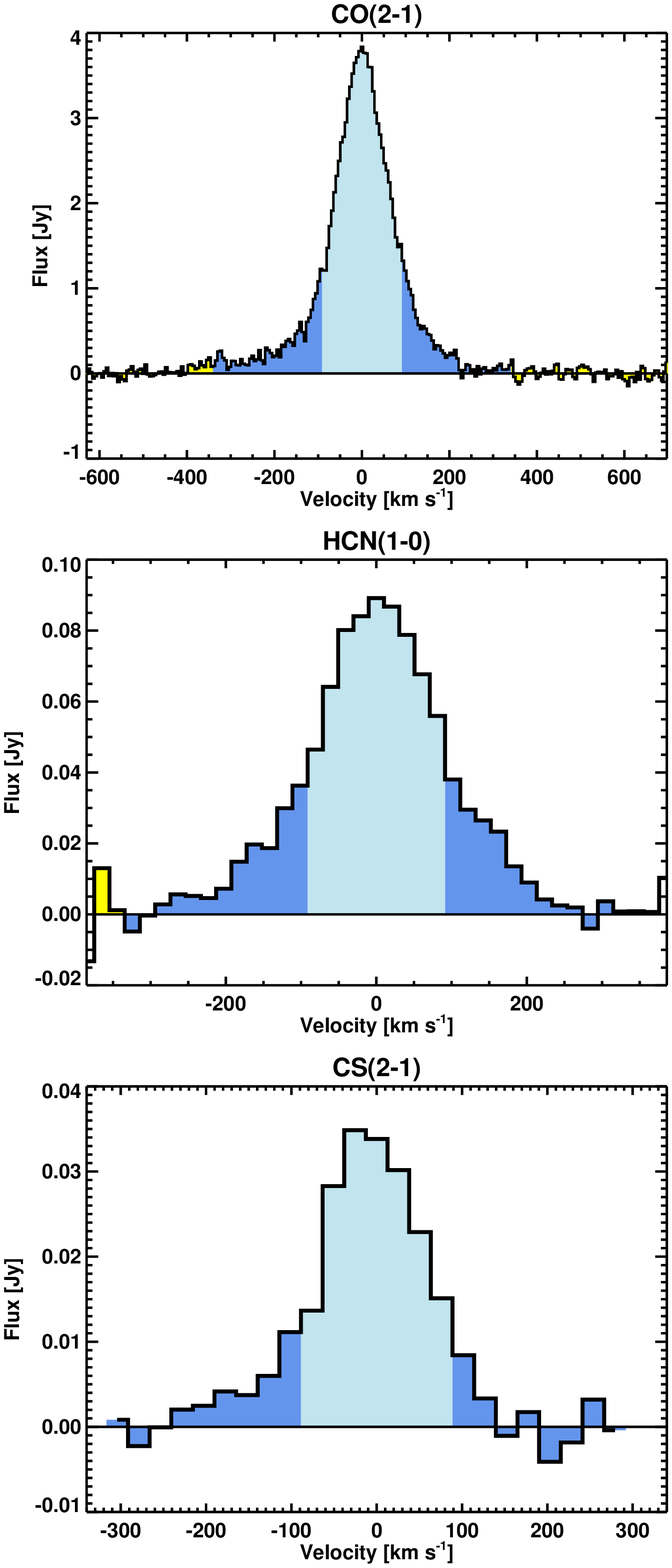}
\caption{Molecular spectra of NGC\,1266 from CARMA and the IRAM 30m. The yellow shading represents emission exceeding the escape velocity ($v_{\rm esc}=364$~km~s$^{-1}$), dark blue shading for emission exceeding the maximal stellar rotation velocity determined by SAURON ($v_{\rm rot} = 92.4$~km~s$^{-1}$; \citealt{krajnovic+12}).  {\bf(Top):} Original CO(2--1) spectrum from \citet{alatalo+11} from the IRAM 30m. {\bf(Middle):} The HCN(1--0) emission from CARMA shows broad wings in both the redshifted and blueshifted directions, beyond the maximal stellar rotation velocity.  This shows that the molecular outflow in NGC\,1266 contains dense gas. {\bf(Bottom):} CS(2--1) from CARMA also shows strong signs of a blueshifted line wing, which confirms that dense gas is taking part in the molecular outflow in NGC\,1266.}
\label{fig:dense_wings}
\end{figure}

\section{The mass and mass flux of the molecular outflow}
\label{sec:massoutflow}
Placing independent constraints on the mass of the outflowing material is essential to understanding the nature of the molecular outflow that exists in NGC\,1266.  Calculating the mass of the outflow in NGC\,1266 has many uncertainties.  When calculating the mass from the CO emission in \citet{alatalo+11}, the $L$(CO)--to--H$_2$ conversion factor is important, and harbors many uncertainties.  An optically thin $L$(CO)--to--H$_2$ conversion factor assumes that the observer is seeing all CO molecules in a given object.  This has the effect of creating the smallest possible conversion to molecular gas mass, assuming solar metallicity to determine the CO abundance \citep{knapp+jura76}.  In most giant molecular clouds in the Milky Way however, the molecular gas is optically thick (i.e. the CO is self-shielding), meaning that we are not seeing all CO emission, therefore there is a larger mass of H$_2$ per observed CO molecule \citep{solomon+87}.  \citet{alatalo+11} used the optically thin $L$(CO)--to--H$_2$ conversion when calculating the mass of the molecular gas in the outflow, in order to obtain the lower limit to the molecular gas.  If the molecular gas in the outflow is not diffuse, then the optically thin assumption is not valid.

The presence of wings in the HCN and CS spectra from Figure \ref{fig:dense_wings} indicate that the molecular outflow is dense and optically thick, therefore requiring that the $L$(CO)--to--H$_2$ conversion is optically thick rather than thin. This means that the original estimate of the molecular outflow mass, $M_{\rm out} = 2.4\times10^7~M_\odot$ \citep{alatalo+11}, is too low, by about an order of magnitude. This also revises upward the mass outflow rate, originally reported to be $13~M_\odot$~yr$^{-1}$.  Using a conversion factor derived from Ultraluminous Infrared Galaxies (ULIRGs; \citealt{sanders+88,downes+98}), with $\alpha_{\rm CO} \approx 1$ using $M_{\rm mol} = \alpha_{\rm CO} L_{\rm CO}$ \citep{bolatto+13}, and the original measurement of the broad wing emission in \citet{alatalo+11} of $I_{\rm CO}$ = 45.4~Jy~km~s$^{-1}$, the updated molecular mass of the molecular outflow is $2.0\times10^8$~M$_\odot$, a factor of $\approx8$ larger than the optically thin conversion factor derivation.  This also up-converts the molecular mass outflow rate, $\dot{M}_{\rm out}$ to $110~M_\odot$~yr$^{-1}$.

The mass outflow rate is now $\gtrsim 100$ times larger than the SFR, which likely rules out the possibility that the mass driving associated with the molecular outflow is due to SF based on applying the rule of thumb from \citet{murray+05}, solidifying the result of \citet{alatalo+11} that the AGN must be the driving source of the molecular outflow.  The momentum ratio $\zeta = \dot{M}v/(L_{\rm IR}/c)$, which needs to be of order unity to use photon momentum driving to explain the mass outflow rate, is also very high. In NGC\,1266, with $L_{\rm IR} \approx 5.2\times10^{43}$~ergs\,s$^{-1}$, $\dot{M} = 110~M_\odot$~yr$^{-1}$, $v = v_{\rm outflow} = 177$~km~s$^{-1}$, $\zeta \approx 65$. Models to explain such high ratios typically invoke energy driven flows \citep{faucher-giguere+12} or require a high optical depth to the AGN, even in the IR \citep{thompson+14}. While both of these phenomena might be present in this system, it is unlikely that they are able to compensate for the nearly two orders of magnitude factor of $\zeta$. \citet{nyland+13} were able to show that momentum driving by coupling to the radio jet would also be sufficient to drive the outflow, where the radio jet mechanical energy of the AGN does not depend on the AGN also being radiatively powerful.  Therefore, this molecular outflow likely cannot be driven by photons, no matter the source, but the radio jet is energetically capable of doing so.

To investigate how long it takes a galaxy to deplete its molecular gas, the rate at which the mass is outflowing is not the optimal property to investigate, but rather the rate at which the gas is escaping the galaxy.  In NGC\,1266, \citet{alatalo+14} presented strong evidence that the gas remains far longer than would be assumed if one calculated the depletion time via $\tau_{\rm dep} \approx M_{\rm gas}/\dot{M}_{\rm out}$, likely pointing to the fact that gas that does not escape falls back into the center.  As shown in Fig. \ref{fig:dense_wings}, the vast majority of the outflowing molecular gas does not have sufficient speed to be driven completely out of the gravitational potential of NGC\,1266.  If the mass driving is not constant, only 2\% of the molecular gas will be capable of completely escaping the galaxy ($\dot{M}_{\rm esc}\approx 2~M_\odot$~yr$^{-1}$), and the remaining outflowing gas will rain back down onto the center.  This possibly means that the AGN-driven outflow is {\em extending} the life of the molecular gas in the system, by prohibiting the molecular gas from forming stars and suppressing SF.  At the current mass {\em escape} rate, NGC\,1266 will be completely depleted of gas in 2 Gyr (assuming the only avenue for depletion is through outflow, if $\dot{M}_{\rm dep} = \dot{M}_{\rm out}+\dot{M}_{\rm SFR}$, $\tau_{\rm dep}$ is closer to 1\,Gyr), rather than in 85\,Myr, as was originally reported in \citet{alatalo+11}.

\section{Two Possible Scenarios to explain the NGC\,1266 nucleus}
\label{sec:scenarios}
There are two possible scenarios that could describe the activity in the center of NGC\,1266, and in particular, the far-IR emission.  Each scenario comes with a set of evidence as well as caveats.  Firstly, NGC\,1266 could have a radiatively powerful, accreting supermassive black hole\footnote{NGC\,1266 does contain a mechanically powerful AGN, evidenced by the radio jet and the energetics of the powerful molecular outflow (\citealt{alatalo+11}; \S\ref{sec:massoutflow})}.   \citet{pellegrini+13} showed that NGC\,1266 has molecular line ratios characteristic of ULIRGs (bright high-order CO transitions and abundant H$_2$O) and that the high transitions of CO either must be powered by strong shocks or an AGN, but that these two scenarios are degenerate. An AGN, whether radiatively powerful or intrinsicially weak is buried under sufficient molecular gas to obscure radiation from the mid-IR to the hard X-rays.  The second scenario is that NGC\,1266 does not contain a radiatively powerful AGN, but instead a deeply buried ultra-compact starbursting cluster that is providing the vast majority of the radiation, reprocessed into far-IR emission.

\subsection{Scenario 1: a radiatively powerful Compton-thick AGN}
\label{sec:scenario1}
Several pieces of evidence point to an radiatively powerful AGN in NGC\,1266. \citet{nyland+13} detect a high brightness temperature ($>1.5\times 10^7$ K) core at 1.65~GHz with the VLBA (shown as a cross in Figs. \ref{fig:cs_chans}, \ref{fig:cs+alma}, \ref{fig:hcn_chans} \& \ref{fig:h13cn}), a brightness temperature two orders of magnitude larger than the theoretical upper limit to the thermal emission that a compact starburst is capable of producing \citep{condon92}. While a radio point source does not necessarily imply the presence of an energetically powerful AGN, the radio concentration presented by \citet{nyland+13} strongly supports the presence of an AGN in the system.  The CS(2--1) map was used to derive a line-of-sight gas column density in the direction of the AGN to be $N_{\rm H}\gtrsim 6 \times 10^{24}~{\rm cm^{-2}}$.  This obscuring column is considered Compton thick, capable of burying a substantial AGN and attenuating  X-rays.  \citet{stern+14} were able to use {\em NuSTAR} to show that 2--10keV X-rays could be completely attenuated in powerful quasars given a sufficiently large column.  If a powerful AGN were present in NGC\,1266, the molecular gas column would be sufficient to attenuate the majority of its emission.

In \S\ref{sec:xrayobs}, we were able to show that given the hard X-ray detection from {\em XMM} and {\em Chandra}, combined with the tentative detection of broad Fe\,K$\alpha$ seem to indicate that given a column of $6\times10^{24}$~cm$^{-2}$ would be able to provide all of the total IR luminosity in the system, as AGN light reprocessed into the far-IR.  Given that the obscuring molecular gas is likely clumpy, there are likely lines-of-sight that are slightly less dense, which would allow more X-rays to escape, leading to a smaller AGN luminosity for the modeled X-ray luminosity.

The mid-IR SED of NGC\,1266 is also characteristic of an AGN, with $S_{8.0}/S_{4.5}$ and $S_{5.8}/S_{3.6}$ placing the galaxy within the AGN selection region of \citet{lacy+04}.  The excess emission seen in the mid-IR (between 10--100$\mu$m) from the {\em Spitzer} IRS (green) also indicates that an AGN component is necessary to match the mid-IR SED. Using the \citet{sajina+12} models and deriving the luminosity associated solely with the hot dust, power law component seen in the SED (Fig. \ref{fig:n1266_sed}), we find the mid-IR luminosity of the AGN component to be at least $\gtrsim 4\times10^{42}$ ergs s$^{-1}$.  This estimate does not account for AGN emission re-radiated from the mid-IR to the far-IR, which is non-negligible due to the substantial column.

Figure \ref{fig:cs+alma} shows that the bulk of the far-IR continuum emission is coming from a region within 30\,pc of the nucleus, rather than the more extended dense gas in the system.  This indicates that a compact source, such as the AGN, could be providing most of the dust heating.  If we assume that the central compact continuum source is heated by the AGN, then the compact emission associated with the AGN accounts for 66\% of the total flux density, and the extended emission accounts for 34\% of the total flux density (after a 30\% correction for flux resolved out by the ALMA beam).

\begin{figure*}
\subfigure{\includegraphics[width=0.49\textwidth,clip,trim=1.5cm 0cm 0.7cm 0cm]{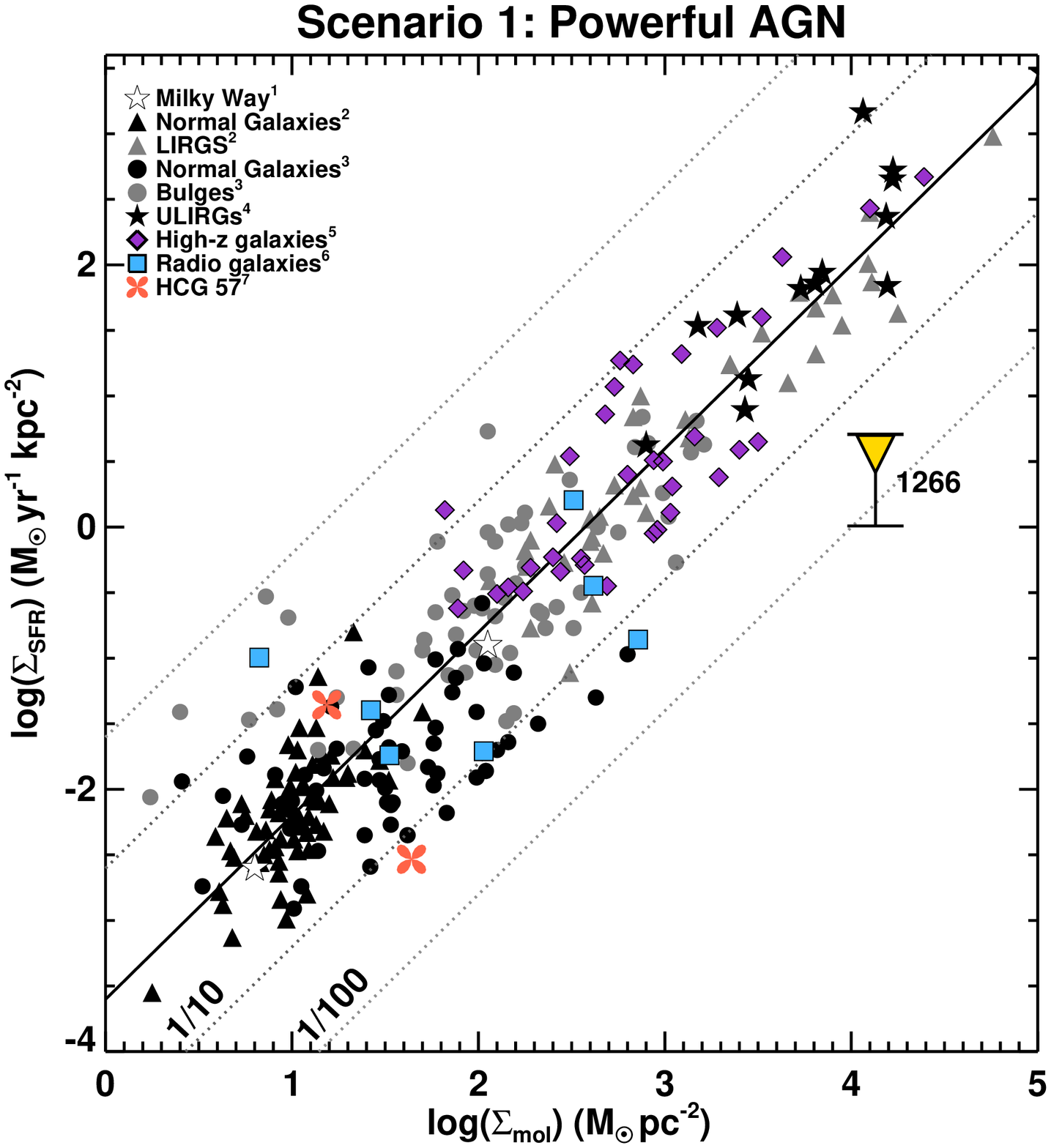}}
\subfigure{\includegraphics[width=0.49\textwidth,clip,trim=1.5cm 0cm 0.7cm 0cm]{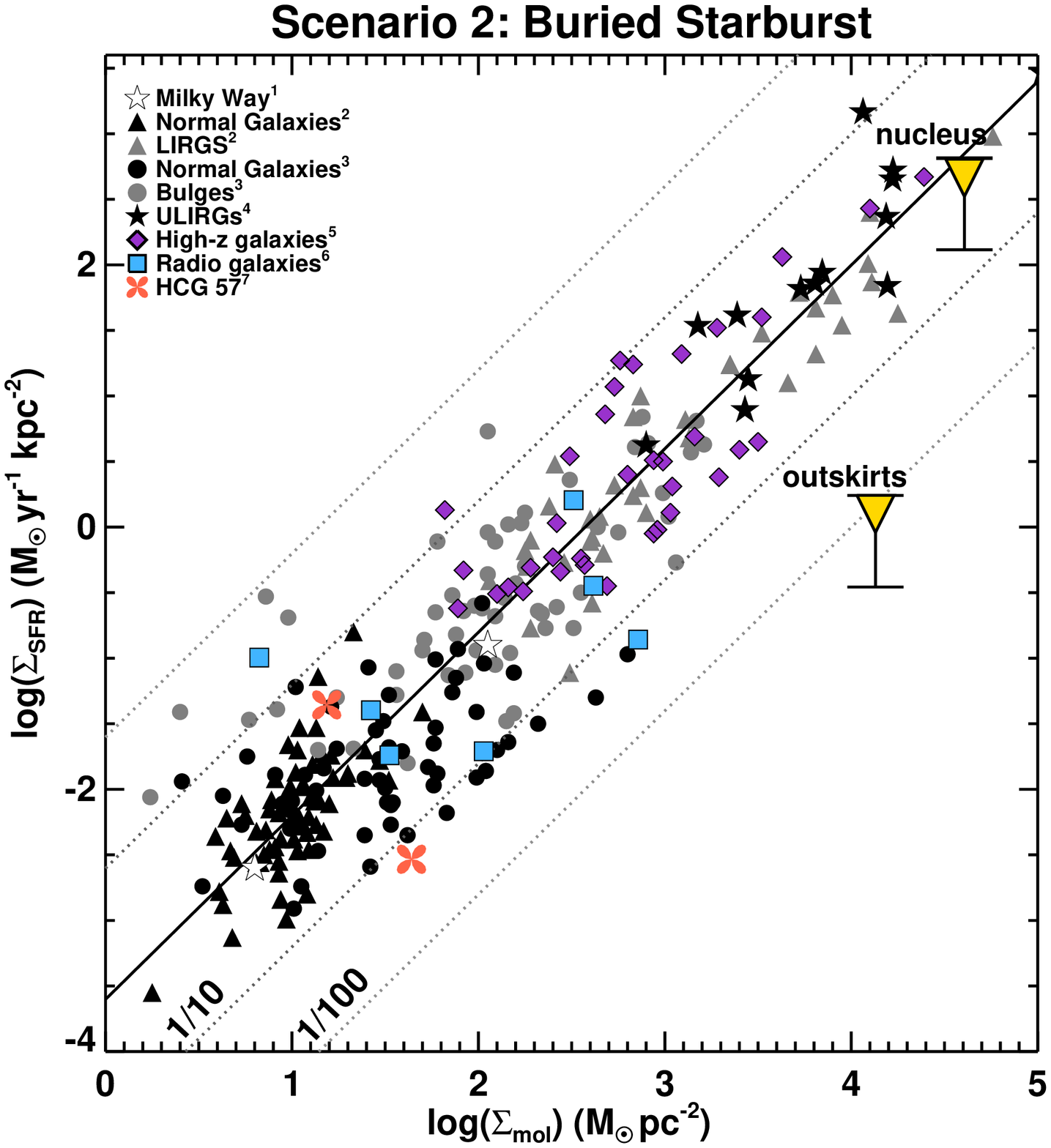}}
\caption{The SFR surface density ($\Sigma_{\rm SFR}$) vs. gas surface density ($\Sigma_{\rm mol}$) of NGC\,1266 is compared to other objects including the Milky Way (1: \citealt{yusef-zadeh+09}), normal star-forming galaxies (2: \citealt{ken98}, 3: \citealt{fisher+13}), ULIRGs (4: \citealt{iono+09}), high redshift objects (5: \citealt{genzel+10}), radio galaxies (6: \citealt{ogle+10}; \S\ref{sec:radgals}) and Hickson Compact Group 57a and 57d (7: \citealt{alatalo+14b}) on the Kennicutt-Schmidt relation, all renormalized to a Salpeter IMF in both panels.  The solid line represents the Kennicutt-Schmidt relation.  The dotted lines represent enhancements/suppressions of the SFR by factors of 10 and 100. {\bf(Left):} If all molecular gas in the NGC\,1266 system is forming stars at the same efficiency (the buried AGN scenario) then the molecular gas is suppressed by a factor of at least 20, though the free-free emission upper limit places that suppression at 50.  {\bf (Right):} Separating the dense gas regions in NGC\,1266 into the nuclear region and the outskirt region (buried starburst scenario) shows that while the nucleus would be forming stars at the proper efficiency, the outskirt region of the dense gas would have to be suppressed by at least a factor of 150.  In either scenario, there is substantial dense gas that does not appear to be forming stars efficiently.}
\label{fig:ks1266}
\end{figure*}

We fit a SED to the central 15$''$ of NGC\,1266, finding a far-IR luminosity of $1.3\times 10^{10} L_{\odot}$ ($5.2\times10^{43}$~ergs\,s$^{-1}$).  Assuming that the far-IR emission from the central source in the 870$\mu$m image is powered by the AGN, the IR decomposition (Figure \ref{fig:n1266_sed}) suggests a total AGN luminosity of $8.6\times10^{9}~L_\odot$ ($3.4\times10^{43}$~ergs\,s$^{-1}$). Assuming NGC\,1266 is currently on the $M-\sigma$ relation \citep{mcconnell+13}, with $\sigma_{1266} \approx 79$~km~s$^{-1}$ \citep{cappellari+13}, the black hole mass would be $\approx 1.7\times10^6$ M$_\odot$. The bolometric AGN luminosity would therefore be about 18\% of the Eddington luminosity, suggesting that NGC\,1266 has energetics similar to a typical Seyfert galaxy \citep{ho2008} if Scenario 1 is correct.


\subsection{Scenario 2: an ultra-compact starburst}

The presence of an AGN in NGC\,1266 does not necessarily confirm the presence of a radiatively powerful AGN. It has been shown that many low-luminosity AGNs also exhibit Fe~K$\alpha$ emission \citep{nandra+07}.  Although the molecular data confirm the Compton-thick line-of-sight column toward the AGN, we cannot rule out the possibility that preferential lines-of-sight in a highly clumpy ISM might allow a larger amount of the X-ray emission from a weak AGN to escape, meaning the source of the far-IR luminosity in the compact core is not definitively due to a radiatively powerful AGN.

It is theoretically possible for a ultra-compact starburst to produce the observed compact far-IR emission \citep{soifer+00,younger+08,walter+09,hopkins+10}.  The total IR luminosity in the compact core is $8.5\times10^9~L_\odot$, with a radius $R<22$\,pc, has a total luminosity density of $\Sigma_{L_{\rm core}} > 5.6\times10^{12}~L_\odot$\,kpc$^{-2}$, well below the maximum luminosity density of star-forming cores (10$^{14}~L_\odot$\,kpc$^{-2}$; \citealt{soifer+00}).  Assuming that this luminosity is primarily produced by O-stars ($M_\star>20~M_\odot$), with light-to-mass ratios of $\approx2300~L_\odot/M_\odot$ \citep{evostars}, the O-star mass in the compact region must be at least $3.7\times10^6~M_\odot$ (or 8\% of the total enclosed mass in the central region, calculated in \S\ref{sec:almaobs}) to sustain the luminosity observed.

The enclosed mass includes the supermassive black hole within, molecular gas, as well as stars.  Using a Salpeter IMF \citep{salpeter55}, the expected total mass fraction of stars with masses $M_\star>20~M_\odot$ should account for $\approx1$\% of the {\em stellar} mass.  The sustainability of this starburst scenario depends on a top-heavy IMF to account simultaneously for the enclosed mass and the large luminosity.  Although unusual, we can not rule out that such a top-heavy IMF might exist in the center of NGC~1266, given the likely presence of a top-heavy IMF in the center of the Milky Way \citep{maness+07,bartko+09}.  While this ultracompact region might host 66\% of the SF in the system, it only hosts $\approx2$\% of the dense molecular gas.  This means that the minority of molecular gas is that which is hosting the vast majority of SF in this scenario.

\subsection{Distinguishing between an AGN and an ultra-compact starburst}
While the current data are able to put strong constraints on the size of the emitting ultra-compact core and mass enclosed, they are not able to definitively identify the source of the radiation.  We need to measure emission coming directly from a Compton-thick AGN, constrain the size of the emitting region such that it is physically impossible for an ultra-compact starburst to be the source or find an alternative measure of SF.  Measuring a hard X-ray flux from {\em NuSTAR} observations should both be able to pin down the the average attenuating column along the line-of-sight to the AGN.  Alternatively, if we are able to show that the luminosity density within the emitting region exceeds $10^{14}~L_\odot$\,yr$^{-1}$, we know it must be an AGN.  If observations are able to show that the radius of the emitting region is smaller than 5.2\,pc ($d<10.4$\,pc or 0.07$''$), then we can confirm that an AGN must be the source of the emission.  The new capabilities of ALMA will be able to reach this resolution, and provide the definitive evidence of the presence of the AGN.  Finally, one might be able to count individual supernova remnants (SNRs) in NGC\,1266 (similar to what has been done in Arp\,220; \citealt{parra+07}) using transatlantic very long baseline interferometry, which should be able to distinguish between a single point source (Scenario 1: the AGN) and multiple SNRs (Scenario 2: ultra-compact starburst).


\begin{figure*}[t!]
\includegraphics[width=6.5in]{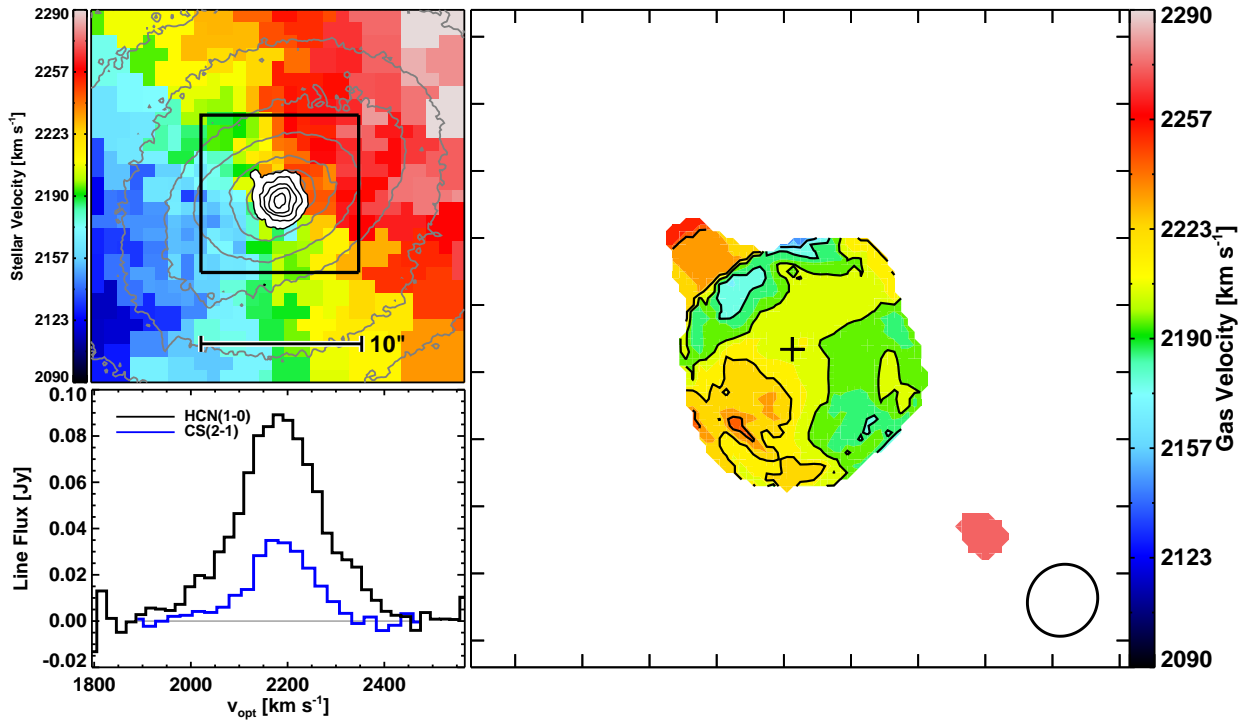}
\caption{{\bf(Top left):} The stellar kinematic map of NGC\,1266 from SAURON \citep{krajnovic+12} is overlaid with the CS(2--1) moment0 map (black contours with white base contour) and the HST {\em Y}-band image (gray contours), showing the stars in NGC\,1266 follow a regular rotation.  The CS(2--1) mean velocity map ({\bf right}). The mean velocity map lacks any distinct pattern of rotation, and certainly very little in the plane of the larger scale stellar rotation, meaning that it is unlikely that the dense gas in NGC\,1266 is rotationally supported.  The CARMA beam is shown on the bottom left of the image, indicating that there are $\sim10$ resolution elements across the CS(2--1) map. {\bf(Bottom left):} Both the HCN(1--0) (black) and CS(2--1) spectra from CARMA are shown, indicating that the velocity width of the dense gas (traced by both CS and HCN) exceeds the rotation velocity seen in the stellar rotation.  It is of note that both the HCN and CS also include broad line wings.}
\label{fig:cs}
\end{figure*}

\section{SF suppression in NGC\,1266}
\label{sec:sfsupp}

Our observations show that the molecular gas is distributed in a configuration with diameter 3.8$''$ (550\,pc) in diameter (Fig. \ref{fig:cs+alma}), with a surface density measured from the resolved average CS(2--1) emission to be $\Sigma_{\rm gas} \approx 10^4~M_\odot$~pc$^{-2}$. Such an extreme gas surface density is more comparable to those found in ULIRGs than to normal SF galaxies. The continuum emission consists of a slightly resolved 870$\mu$m\ compact source (\S\ref{sec:almaobs}; Figure \ref{fig:cs+alma}, top) with a radius $\approx$\,22\,pc, surrounded by a halo of lower surface brightness emission visible in the 870$\mu$m\ radial profile (Figure \ref{fig:cs+alma}, bottom), cospatial with the molecular gas seen in the CS(2--1) emission. In the compact core emission, the virial estimate of the mass from H$^{13}$CN detection as well as CS(2--1) emission was 3 times larger than the integrated gas surface density, or $\Sigma_{\rm dense,nuc}\approx3\times10^4$\,M$_\odot$\,pc$^{-2}$.

In \S\ref{sec:sf}, we estimated the SFR in NGC\,1266 from the free-free emission, we are able to obtain a limit on the SFR of $\approx0.87~M_{\odot}~{\rm yr^{-1}}$, with an upper limit determined by the [Ne\,{\sc ii}] emission of 1.5\,$M_\odot$\,yr$^{-1}$.  Figure \ref{fig:ks1266} shows the mass surface density vs. SFR surface density of NGC\,1266 in both scenarios, compared to other star-forming objects including the Milky Way \citep{yusef-zadeh+09}, normal star-forming galaxies \citep{ken98,fisher+13}, ULIRGs \citep{iono+09}, high redshift objects \citep{genzel+10}, radio galaxies \citep{ogle+10} and HCG\,57 \citep{alatalo+14b}, on the Kennicutt-Schmidt (K-S) relation \citep{ken98}. 

In the case of the powerful imbedded AGN (Scenario 1), the SF is distributed across the entire dense molecular core.  In the dense gas region, $\Sigma_{\rm H_2} \approx 10^4$\,$M_\odot$\,yr$^{-1}$.  The low SFR surface density places NGC\,1266 at least a factor of 50 below the K-S relation.  In the case of an imbedded ultra-compact starburst (Scenario 2), the SF is distributed between the compact (66\%) and the outlying (34\%) regions, leading to SFR$_{\rm core} = 1$\,$M_\odot$\,yr$^{-1}$ and SFR$_{\rm outskirts} = 0.5$\,$M_\odot$\,yr$^{-1}$, and $\Sigma_{\rm H_2, core} \approx 3\times10^4$\,$M_\odot$\,yr$^{-1}$ (and only contains 2\% of the dense molecular gas) and $\Sigma_{\rm H_2, outskirts} \approx 10^4$\,$M_\odot$\,yr$^{-1}$.  Figure \ref{fig:ks1266}, Scenario 2 shows that while the nuclear region of NGC~1266 sits on the K-S relation (at the upper right corner, with the ULIRGs), the outskirt region is suppressed by at least a factor of 150.  In both scenarios, the SF in some subset of the dense molecular gas in NGC~1266 is suppressed.

\subsection{Driving SF suppression}
In both scenarios, there is a large component of molecular gas that is not forming stars at normal efficiency.  The molecular gas in the majority of star-forming systems is thought to be in a rotationally supported disk.  Figure \ref{fig:cs} demonstrates that there is a severe kinematic mismatch between the stellar kinematics and the dense gas kinematics, implying that the dense gas is not rotationally supported.  In order for SF to be suppressed, some other force must balance gravitational collapse.  In normal galaxy discs, the parameter used to describe this balance is the Toomre $Q$ parameter \citep{toomreQ}:
 \[Q = \frac{\sigma\kappa}{3 G \Sigma}\]
 
\noindent where $\sigma$ is the gas velocity linewidth, $\kappa$ is the epicyclic frequency, $G$ is the gravitational constant, and $\Sigma$ is the gas surface density.  For NGC\,1266, $\kappa \approx 1.8$~km~s$^{-1}$~pc$^{-1}$ \citep{alatalo+11}, $\sigma \approx 30$~km~s$^{-1}$ \citep{pellegrini+13}, and $\Sigma_{\rm H_2} \approx 10^4~M_\odot$~pc$^{-2}$.  Molecular disks start collapsing when $Q\lesssim1$.  Using the parameters above, $Q\approx0.4$, which means that NGC\,1266 should be very unstable against gravitational collapse and therefore forming stars prolifically, which even the upper limit of SF is unable to support.  Fig. \ref{fig:cs} supports an alternative explanation.  The high velocity linewidth seen in the spectrum of the CS gas, the disordered motion observed both in the channel maps as well as the mean velocity map do not represent a face-on disc.  In fact, the dense gas emission does not show the same velocity gradient as is observed in the CO(1--0) emission from \citet{alatalo+11}.  Instead, it is possible that the bulk of the dense gas (which is closest to the AGN) is distributed spherically.

In this case, there is an alternative way to balance gravity, and hinder collapse, namely turbulence injected by the AGN outflow.  In Scenario 1 (AGN), the molecular outflow is likely to be able to provide a source of that turbulent injection into the molecular gas as a whole.  In Scenario 2 (starburst), it might be feedback that exists between the dense molecular gas and both the molecular outflow as well as the winds from the ultra-compact starbursting region. Both of these phenomena inject further turbulence into the outskirt molecular gas, leading to the observed suppression.

To obtain a rough estimate of the required outflow velocities, we assume an ordered radial outflow, $v_{\rm rad}$, as the dominant component of the kinetic energy. The necessary radial velocity is then the free-fall velocity for a uniform density sphere:
\[v_{\rm rad} = \sqrt{\frac{3GM_{\rm enc}}{5R}}\]

For NGC\,1266, with M$_{\rm enc} \approx $ M$_{\rm H_2} \approx 10^9$ M$_{\odot}$ \citep{alatalo+11} and $R = 275$ pc, the required radial velocity is $v_{\rm rad}$ = 100 km s$^{-1}$.  Both the CO narrow-band FWHM as well as the FWHM in the spectrum of CS (Figure \ref{fig:cs}) are $\sim$ 100~km~s$^{-1}$.  Any bulk motion of this kind would quickly become turbulent \citep{appleton+06,appleton+13}, which could suppress SF until all the kinetic energy (E$_{\rm kin}$ = 1/2 M$_{\rm H_2} \sigma^2 = 2.4\times10^{55}$~ergs) was dissipated on a timescale $\tau=\Lambda/\sigma\approx5.5$ Myr, where $\Lambda$ is the characteristic scale of the turbulence (assumed 275~pc) and $\sigma_{\rm gas}$ = 48.5 km s$^{-1}$.  Indeed, turbulent energy dissipation is consistent with the large warm H$_2$ luminosity for NGC\,1266 (L$_{\rm H_2,warm} \approx 8\times10^{40}$~ergs\,s$^{-1}$; \citealt{roussel+07,alatalo+11}).  If our explanation is correct, then dissipation of turbulent energy associated with the outflow (E$_{\rm kin}/\tau=1.3\times10^{41}$~ergs\,s$^{-1}$) is capable of energizing a large fraction of the H$_2$ line luminosity of the warm gas.  In order to maintain the activity and avoid the gas collapsing to the center in a free-fall time of $t_{\rm ff} = (\frac{4\pi}{5}G\rho)^{-1/2}=2.8$~Myr, the turbulence must be maintained by constant energy input. This mechanism has been suggested to explain infertile
gas in radio galaxies \citep{nesvadba+10,guillard+14}, Stephan's Quintet \citep{appleton+06,guillard+12} and NGC\,1266.  In fact, a re-analysis of the warm H$_2$ emission in the {\em Spitzer} IRS spectrum (taking into account extinction using the 9.7$\mu$m silicate feature, confirmed that NGC\,1266 is a Molecular Hydrogen Emission Galaxy (MOHEG; \citealt{ogle+06}), with $L$(H$_2$, warm)/$L$(7.7$\mu$m PAH) $>0.1$.

If the driving is due to a radio jet \citep{nesvadba+10,ogle+10,guillard+12,guillard+14}, then the support of the gas against gravitational collapse can last at most a few$\times10^7$~yr per episode of AGN activity. This could be prolonged as the  vast majority (98\%) of the cold molecular gas does not have speeds exceeding the escape velocity of the galaxy \citep{alatalo+11}.  This gas will thus rain back onto the nucleus, re-impart its kinetic energy into the molecular core, providing longer-timescale energy injection to counterbalance gravitational collapse.  If the gas in NGC\,1266 were forming stars at normal efficiency, the depletion time would be $\approx100$~Myr, five times too short, given the present stellar population \citep{alatalo+14}.  If the current SF suppression remains consistent into the future, then the molecular gas could remain in the nucleus as long as 7~Gyr, allowing the molecular gas to remain in contact with the AGN for much longer than has been hypothesized \citep{cicone+13}.

\subsection{SF suppression as a method for maintaining the M-$\sigma$ relation}
We propose that NGC\,1266 contains an AGN whose black hole is being built up to bring it onto the $M-\sigma$  relation following a minor merger event 500 Myr ago. We hypothesize that prior to this minor merger, the supermassive black hole within NGC\,1266 was on the $M-\sigma$ relation.  This merger event triggered a burst of SF that increased the mass of the bulge by about 10\% ($3\times 10^8 M_{\odot}$; \citealt{alatalo+14}) and left a remnant  $\approx 2\times 10^9 M_{\odot}$  of gas which failed to form stars in the initial burst. The corresponding growth in the black hole mass may have involved multiple cycles of AGN activity over the past 500 Myr that heavily suppress further SF in the nucleus by inducing turbulence in the ISM, but which will also eventually deprive the AGN of fuel.  Our results are currently inconclusive as to whether the AGN is currently in a high accretion rate phase of its cycle (Scenario 1), or a low accretion-rate phase (Scenario 2). In either case the large amount of gas in the nuclear region suggests that the cycles will continue until the black hole becomes massive enough that its Eddington luminosity rises to the point that the outflow is able to finally expel the remaining gas. At this point, the black hole mass will match that expected on the $M-\sigma$ relation.

\section{Conclusions}
\label{conc}

We have presented new CARMA and ALMA observations of the dense molecular gas and continuum of NGC\,1266, which have updated properties of the mass outflow and SFR of this AGN-driven molecular outflow system. Using the radio free-free emission present in the 1mm band of ALMA, we have made the most accurate estimate of the SFR in this system to date, 0.87$M_\odot$~yr$^{-1}$ (with approximately a factor of 2 uncertainty). The [Ne\,{\sc ii}]-derived SFR provide an upper limit to the SFR of 1.5\,$M_\odot$\,yr$^{-1}$.


Observations of HCN(1--0), CS(2--1) and H$^{13}$CN(3--2) confirm the presence of broad line wings in the dense gas, requiring an upward revision of the mass outflow rate originally quoted in \citet{alatalo+11} from $13~M_\odot$~yr$^{-1}$ to $\approx110~M_\odot$~yr$^{-1}$, because an optically thin $L$(CO)--to--H$_2$  conversion factor did not apply. Despite this, only $\approx 2 M_\odot$~yr$^{-1}$ is escaping the galaxy as molecular gas. Our higher mass outflow rate suggests that the driving mechanism responsible for this high momentum ratio outflow is likely momentum-coupling to the radio jet from the AGN, rather than to photons from either the AGN or a starburst.

There are two possible scenarios that explain the activity in the nucleus of NGC\,1266.  Scenario 1, supported by the tentative detection of Fe\,K$\alpha$ and the Compton-thick column of molecular gas in the line-of-sight to a VLBA detection, is that the compact 870$\mu$m emission is due to a deeply buried, powerful AGN.  Scenario 2, in which an ultra-compact starburst, buried deeply under the molecular gas provides the far-IR emission, though requires a top-heavy IMF to sustain.  In either scenario, the SF in the majority of the dense molecular gas is suppressed by between 50--150.

The most likely explanation for the suppression of SF in the dense gas is that turbulence injected by, or during the formation of, the molecular outflow has been able to balance the gravitational potential of the dense gas in the center, and thus turbulent motions are a reasonable explanation for the SF suppression seen.

Our detailed study of NGC\,1266 provides the first example of an intermediate mass galaxy presenting evidence of SF regulation by an AGN. We further hypthesise that it is an example of an intermediate mass galaxy building its black hole mass onto the $M-\sigma$ relation. As the black hole mass increases, the maximum AGN luminosity will increase, to the point where the outflow may be strong enough to expel most of the gas from the system.

\newpage
\acknowledgments
K.A. thanks Daniel Perley for the Keck time that provided the NGC\,1266 $u'$ data.  K.A. also thanks Chris McKee, Carl Heiles, Kartik Sheth, Adam Leroy, Mark Krumholz and Nathan Roth for insightful discussions.  The authors would also like to thank the anonymous referee for useful recommendations, as well as pointing us to the publically available {\em XMM} data, which we were able to use to further clarify the activity present in this system.
K.A. is supported by funding through Herschel, a European Space Agency Cornerstone Mission with significant participation by NASA, through an award issued by JPL/Caltech.  The research of K.A. was also supported by the NSF grant AST-0838258.  K.N. and D.S.M. are supported by NSF grant 1109803.  SLC was supported by ALMA-CONICYT program 31110020.  PC gratefully acknowledges support  from the NASA ATP program through NASA grant NNX13AH43G, and NSF grant AST-1255469.

This paper makes use of the following ALMA data: ADS/JAO.ALMA\#2011.0.00511.S. ALMA is a partnership of ESO (representing its member states), NSF (USA) and NINS (Japan), together with NRC (Canada) and NSC and ASIAA (Taiwan), in cooperation with the Republic of Chile. The Joint ALMA Observatory is operated by ESO, AUI/NRAO and NAOJ.  The National Radio Astronomy Observatory is a facility of the National Science Foundation operated under cooperative agreement by Associated Universities, Inc.  Support for CARMA construction was derived from the states of California, Illinois, and Maryland, the James S. McDonnell Foundation, the Gordon and Betty Moore Foundation, the Kenneth T. and Eileen L. Norris Foundation, the University of Chicago, the Associates of the California Institute of Technology, and the National Science Foundation. Ongoing CARMA development and operations are supported by the National Science Foundation under a cooperative agreement, and by the CARMA partner universities. This research has made use of NASA's Astrophysics Data System. Based in part on observations made with the NASA Galaxy Evolution Explorer.  GALEX is operated for NASA by the California Institute of Technology under NASA contract NAS5-98034."

\appendix
\section{Radio galaxy points on the Kennicutt-Schmidt relation}
\label{sec:radgals}

For the radio galaxies presented in Figure \ref{fig:ks1266}, we updated the original work by \citet{nesvadba+10}, who used the 7.7$\mu$m PAH-derived SFR from \citet{ogle+10}, and assumed a radius of 2.5~kpc for all radio galaxies without interferometric CO maps.  We have determined the SFRs for these radio galaxies using the {\em Spitzer} Multiband Imaging Photometer (MIPS) 70$\mu$m images, and measured the total flux within a given aperture, then converted the $70\mu$m flux (assuming all flux was from SF) to a SFR using \citet{calzetti+10}.   We have only included those radio galaxies which had definitively determined radii, either from CO measurements (3C326N; \citealt{nesvadba+10}; 3C293; \citealt{evans+98}; 3C31; \citealt{okuda+05}) or from the 8$\mu$m PAH extent derived from {\em Spitzer} IR Array Camera (IRAC) images \citep{irac}, determined by subtracting the convolved 3.6$\mu$m emission with a scale factor of 0.26 \citep{shapiro+10} from the $8\mu$m {\em Spitzer} data.   These updated radio galaxies on average are more suppressed than the normal SF galaxies, but not as dramatically suppressed as NGC\,1266.  In NGC\,1266, all molecular gas has been evacuated except in the very center \citep{alatalo+11}.  On the other hand, the CO and SF within radio galaxies are integrated measurements, and include the entire galaxy.  It is likely that the same mechanisms are present, but the AGN's dominance does not apply to the whole of the molecular gas in the radio galaxy systems, as it does in NGC\,1266.

\end{document}